\documentclass[twocolumn, times, tighten,twocolappendix]{aastex63}

\usepackage{graphicx,multirow,hyperref,url,color,xspace}
\usepackage{amsmath,amssymb}

\newcommand{\msun}{{\rm M}_{\sun}}

\newcommand{\nustar}{{\textit{NuSTAR}}\xspace}
\newcommand{\nicer}{NICER\xspace}
\newcommand{\integral}{{\textit{INTEGRAL}}\xspace}
\newcommand{\source}{{MAXI J1820+070}\xspace}

\newcommand{\appropto}{\mathrel{\vcenter{
  \offinterlineskip\halign{\hfil$##$\cr
    \propto\cr\noalign{\kern2pt}\sim\cr\noalign{\kern-2pt}}}}}

\defcitealias{Zdziarski21b}{Z21}
\newcommand{\zdz}{{\citetalias{Zdziarski21b}}\xspace}

\begin{document}

\title{Hybrid Comptonization and Electron-Positron Pair Production in the Black-Hole X-Ray Binary MAXI J1820+070}
\shorttitle{Hybrid Comptonization and e$^\pm$ Pair Production}

\author{Andrzej A. Zdziarski}
\affiliation{Nicolaus Copernicus Astronomical Center, Polish Academy of Sciences, Bartycka 18, PL-00-716 Warszawa, Poland; \href{mailto:aaz@camk.edu.pl}{aaz@camk.edu.pl}}
\author{Elisabeth Jourdain}
\affiliation{CNRS; IRAP; 9 Av. colonel Roche, BP 44346, F-31028 Toulouse cedex 4, France}
\affiliation{Universit{\'e} de Toulouse; UPS-OMP; IRAP; Toulouse, France}
\author{Piotr Lubi\'nski}
\affiliation{Institute of Physics, University of Zielona G\'{o}ra, Licealna 9, 
PL-65-417 Zielona G\'{o}ra, Poland}
\author{Micha{\l} Szanecki}
\affiliation{Faculty of Physics and Applied Informatics, {\L}{\'o}d{\'z} University, Pomorska 149/153, PL-90-236 {\L}{\'o}d{\'z}, Poland}
\author{Andrzej Nied{\'z}wiecki}
\affiliation{Faculty of Physics and Applied Informatics, {\L}{\'o}d{\'z} University, Pomorska 149/153, PL-90-236 {\L}{\'o}d{\'z}, Poland}
\author{Alexandra Veledina}
\affiliation{Department of Physics and Astronomy, FI-20014 University of Turku, Finland}
\affiliation{Nordita, KTH Royal Institute of Technology and Stockholm University, Roslagstullsbacken 23, SE-10691 Stockholm, Sweden}
\affiliation{Space Research Institute of the Russian Academy of Sciences, Profsoyuznaya Str. 84/32, 117997 Moscow, Russia}
\author{Juri Poutanen}
\affiliation{Department of Physics and Astronomy, FI-20014 University of Turku, Finland}
\affiliation{Space Research Institute of the Russian Academy of Sciences, Profsoyuznaya Str. 84/32, 117997 Moscow, Russia}
\affiliation{Nordita, KTH Royal Institute of Technology and Stockholm University, Roslagstullsbacken 23, SE-10691 Stockholm, Sweden}
\author{Marta A. Dzie{\l}ak}
\affiliation{Nicolaus Copernicus Astronomical Center, Polish Academy of Sciences, Bartycka 18, PL-00-716 Warszawa, Poland; \href{mailto:aaz@camk.edu.pl}{aaz@camk.edu.pl}}
\author{Jean-Pierre Roques}
\affiliation{CNRS; IRAP; 9 Av. colonel Roche, BP 44346, F-31028 Toulouse cedex 4, France}
\affiliation{Universit{\'e} de Toulouse; UPS-OMP; IRAP; Toulouse, France}

\shortauthors{Zdziarski et al.}

\begin{abstract}
We study X-ray and soft gamma-ray spectra from the hard state of the accreting black-hole binary MAXI J1820+070. We perform analysis of two joint spectra from \textit{NuSTAR} and \textit{INTEGRAL}, covering the range of 3--650 keV, and of an average joint spectrum over the rise of the hard state, covering the 3--2200 keV range. The spectra are well modelled by Comptonization of soft seed photons. However, the distributions of the scattering electrons are not purely thermal; we find they have substantial high-energy tails, well modelled as power laws. The photon tail in the average spectrum is detected well beyond the threshold for electron-positron pair production, 511 keV. This allows us to calculate the rate of the electron-positron pair production and put a lower limit on the size of the source from pair equilibrium. At the fitted Thomson optical depth of the Comptonizing plasma, the limit is about 4 gravitational radii. If we adopt the sizes estimated by us from the reflection spectroscopy of $>$20 gravitational radii, the fractional pair abundance becomes much less than unity. The low pair abundance is confirmed by the lack of both an annihilation feature and of a pair absorption cutoff above 511 keV in the average spectrum.
\end{abstract}

\section{Introduction}
\label{intro}

Compton scattering of soft seed photons has for long been shown to explain well hard X-ray spectra of the accreting black-hole (BH) X-ray binaries (XRBs) in their hard spectral state. The electron distribution appears to be predominantly thermal at mildly relativistic temperatures, $kT_{\rm e}$, see, e.g., \citet{DGK07} and \citet{Z20_thcomp} and references in those works. However, the form of the spectral high-energy cutoff in some sources indicates the presence of a substantial high-energy non-thermal tail beyond a Maxwellian electron distribution (e.g., \citealt{McConnell02, Wardzinski02, PV09, ZMC17, Walter17, Cangemi21}). On the other hand, instead of being thermal, the dominant Comptonization process could be scattering on the bulk motion of fast-moving plasmoids generated in magnetic reconnection, in which case a photon tail due to electron acceleration can appear as well \citep{Beloborodov17, Sironi20}. 

A related effect is e$^\pm$ pair production in photon-photon collisions. High-energy tails in the spectra of sources explained by hybrid Comptonization usually cross the threshold for this process, $m_{\rm e}c^2$, significantly increasing the number of pair-producing photons with respect to the purely thermal case. Pair equilibria (in which the rate of pair production is balanced by that of pair annihilation) in hybrid plasmas were studied in \citet{ZLM93}, \citet{Coppi99}, \citet{G99}, \citet{McConnell02}, \citet{MB09} and \citet{Fabian17}.  

Here we study these effects in \source. It is a transient BH XRB, whose outburst was discovered in 2018 \citep{Tucker18,Kawamuro18}. The source is relatively nearby, at an accurately measured distance of $d\approx 3.0\pm 0.3$\,kpc \citep{Atri20}.  Also, the inclinations of both the binary and its radio jet are well constrained, as $i_{\rm b}\approx 66\degr$--$81\degr$ \citep{Torres20}, $i_{\rm j}\approx 63\degr\pm 3\degr$ \citep{Atri20}, respectively. The BH mass is given by $M\approx (5.95\pm 0.22)\msun/\sin^3 i_{\rm b}$ \citep{Torres20}. The outburst was extensively monitored by a number of observatories, in particular by {\it Nuclear Spectroscopic Telescope Array} (\nustar; \citealt{Harrison13}), the Spectrometer on \integral (\citealt{Roques03}; SPI), and the Imager on Board the \integral Satellite (\citealt{Ubertini03}; IBIS). Here, we study spectra from contemporaneous observations by those instruments during the initial hard state of the outburst. They provide high-quality broad-band spectra in the range from 3\,keV up to $\gtrsim$1\,MeV. 

\section{Observations and data reduction}
\label{data}

\renewcommand{\arraystretch}{0.9}\begin{table*}\centering
\caption{Contemporaneous observations of \source with \integral and \nustar in the hard state  
}
\vskip -0.4cm                               
\begin{tabular}{lcccccccccc}
\hline
Epoch & \integral & SPI Start time & Exposure  & IBIS Start time & Exp.\ ISGRI &\nustar Obs. ID & Start time & Exposure A \\
& revolution& End time && End time &Exp.\ PICSiT && End time&  Exposure B \\
\hline
1 & 1934 &58201.555& 13604 &58201.544 &8796 &90401309008 &58201.526&3046  \\
&&58201.757&&58201.757&&&58201.766 &3214\\
2 & 1938 &58212.181 &28507 &58212.181 &8305 &90401309012 &58212.200&12334   \\
&&58212.606&&58212.393&&&58213.177& 12964\\
A &1931--1951 &58193.453& 1147756 &58193.455 &765345& 90401309004--16 & 58198.036& 50176\\
&&58246.894&&58246.892 & 1156304 &&58242.643& 52799\\
\hline
\end{tabular}
\tablecomments{
The times are given in MJD, and the exposures are effective in seconds. The average \nustar spectrum is from 8 observations with the Obs.\ IDs 904013090(04, 06, 08, 10, 12, 13, 14, 16).}
\label{log}
\end{table*}

We use the \integral data from SPI and IBIS. The latter consists of the ISGRI and PICSiT detectors. We use the SPI data as published in \citet{Roques19}. Data reduction and spectral extraction for ISGRI was done using the {\sc osa} v.\ 11.1 software \citep{Courvoisier03}. The analysis of the PICSiT data follows the method of \citet{Lubinski09}.

We also use data from \nustar. They were reduced with {\sc heasoft} v.6.25, the {\tt NuSTARDAS} pipeline v.1.8.0, and {\tt CALDB} v.20200912. We set {\tt saamode=strict}, {\tt tentacle=yes} and {\tt STATUS== b0000xxx00xxxx000}. The source region is a $60''$ circle centered on the peak brightness. We group the data to signal-to-noise ratio $\geq$50, but only below 69\,keV, so we can utilize the 3--79\,keV band to the maximum.

The chosen data sets are listed in Table \ref{log}. Epochs 1 and 2 consist of overlapping \integral and \nustar observations. They correspond to the beginning of a plateau phase on the count-rate/hardness plot, see fig.\ 2 in \citet{Buisson19}. For them, the PICSiT data have low statistics and we do not use them. Then, we analyze the average \integral spectra for the initial hard state, over the span of 53 d, and the corresponding average \nustar spectra, which set is denoted as A in Table \ref{log}. These \integral observations start in the middle of the initial sharp rise phase of the outburst and continue through the middle of the plateau phase. This data set includes the average PICSiT spectrum.

We use the SPI spectra from 23\,keV up to 650\,keV for epochs 1, 2, and up to 2.2\,MeV for the average spectrum, and include in them a 0.5\% systematic error (\citealt{Roques19}; added in quadrature). We add the same systematic error to the average \nustar spectra, in order to avoid the broad-band fit to be dominated by the very good statistics of the latter. However, following previous papers on the \nustar data from \source, e.g., \citet{Buisson19}, we do not add a systematic error to the individual spectra. Given apparent calibration inaccuracies in the ISGRI and PICSiT data, we add a 1\% systematic error to each of them. Also, the fluxes from the ISGRI at its lowest energies are substantially below those of the other detectors, and thus we use it in the 32--500\,keV range only. We use the PICSiT data in the 0.24--2\,MeV range.
 
\section{Spectral Fits}
\label{fits}

\subsection{The method}
\label{method}

In our spectral analysis, we follow the treatment of \citet{Zdziarski21b} (\zdz) with some modifications related to the presence of high-energy non-thermal tails in the studied spectra. We use the X-ray fitting package {\sc{xspec}} \citep{Arnaud96}. The fit uncertainties are for 90\% confidence, $\Delta\chi^2 \approx 2.71$. Residual differences between the calibration of the \nustar Focal Plane Modules A and B and the \integral detectors SPI, ISGRI and PICSiT are accounted for by the model {\tt plabs}, which multiplies the model spectra by $K E^{-\Delta\Gamma}$. We set $K$ and $\Delta\Gamma$ fixed at 1 and 0, respectively, for the \nustar A module, and find them fitted to the module B as 0.99 and $\approx +0.01$, respectively. For the \integral epochs 1, 2, we find $\Delta\Gamma\approx +0.008$--0.015 and $K\approx 1.24$--1.28. We account for the ISM absorption using {\tt tbabs} \citep{WAMC00} using the elemental abundances of \citet{AG89}, with the column density toward the source of $N_{\rm H}=1.4\times 10^{21}$\,cm$^{-2}$ (e.g., \citealt{Kajava19}). At this low value, absorption of photons at $E\geq 3$\,keV is weak, but still noticeable.

We model spectra from Comptonization going directly to the observer and those reflected from an accretion disk around a Kerr BH taking into account atomic and relativistic effects. We use the {\tt reflkerr} \citep{Niedzwiecki19} and {\tt reflkerr\_bb} routines. The former assumes the Comptonization is on Maxwellian electrons, using the {\tt compps} code \citep{PS96}. The {\tt reflkerr\_bb} is a new routine\footnote{See \url{users.camk.edu.pl/mitsza/reflkerr}. It can also account for blackbody-like disk emission due to its irradiation and viscous disk dissipation, analogously to the lamppost version {\tt reflkerr\_lpbb} \citep{Zdziarski21}. However, we do not use that feature in this work.}, which allows for the presence of a power-law tail in the distribution, using a modified version of {\tt compps}\footnote{The standard version of {\tt compps} allows for the presence of a tail in the form of ${\rm d}N_{\rm e}/{\rm d}\gamma\propto \gamma^{-p}$ only, which does not correspond to a physical acceleration for low values of $\gamma$. On the other hand, both versions give accurate spectra for any $\tau_{\rm T}$, with the number of calculated scatterings $\approx 50+4 \tau_{\rm T}^2$ (at the expense of a long calculation time for large values of $\tau_{\rm T}$).}. The tail is parametrized by the Lorentz factor ($\gamma_{\rm min}$) at which the electron distribution switches from thermal to a power law in the momentum with an index, $p$, ${\rm d}N_{\rm e}/{\rm d}(\beta\gamma)\propto (\beta\gamma)^{-p}$ (where $\beta c$ is the electron velocity), up to a maximum Lorentz factor, $\gamma_{\rm max}$, assumed to be $10^3$.

The thermal electrons are parametrized by their temperature, $kT_{\rm e}$, and the Compton parameter, $y\equiv 4\tau_{\rm T} kT_{\rm e}/m_{\rm e}c^2$, where $\tau_{\rm T}$ is the Thomson optical depth of the plasma. The Comptonizing cloud is assumed to be spherical and we set ${\tt geom}=0$ in {\tt compps}, which gives a fast method, which, however, overestimates $y$. We therefore use an option of ${\tt geom}=-5$ (for a sinusoidal distribution of the seed photons in a sphere, \citealt{ST80}), which calculation is very slow but accurate, in order to find the actual values of $y$, and thus of $\tau_{\rm T}$. We also evaluate the average 2--10\,keV power-law spectral index, $\Gamma$, of a fitted Comptonization component (unabsorbed and without reflection). The seed photons for Comptonization are assumed to have a blackbody distribution with the temperature of $kT_{\rm bb}=0.2$\,keV, following the results from the Neutron star Interior Composition ExploreR (\nicer; \citealt{Gendreau16}) of \citet{Wang20_HXMT}. 

In {\tt reflkerr}, the reflected spectra in the rest frame are calculated using {\tt xillverCp} (v.\ 1.4.3; \citealt{GK10,Garcia18}) up to $\approx$10\,keV and {\tt ireflect} \citep{MZ95} at higher energies, see \citet{Niedzwiecki19} for details. Both the direct and reflected components are integrated over the surface of a Keplerian disk taking into account the relativistic effects. Here, we assume the standard emissivity profile of $\propto R^{-3}$, which is the same as the disk viscous dissipation at $R\gg R_{\rm ISCO}$, where $R_{\rm ISCO}$ is the radius of the innermost stable circular orbit. The reflection fraction, ${\cal R}$, is defined in {\tt reflkerr} as the ratio of the flux irradiating the disk to that emitted outside in a local frame. We assume a rotating BH with the dimensionless spin of $a_*=0.998$, for which $R_{\rm{ISCO}}\approx 1.237 R_{\rm{g}}$, where $R_{\rm g}\equiv GM/c^2$. However, the effect of this assumption is negligible at $R\gg R_{\rm{ISCO}}$.

We then use the same spectral model as \zdz, with two Comptonization components, soft and hard, which reflect from two zones in the disk. As found in \zdz, and confirmed in this study for the individual observations, fits with this model show that the hard component dominates both the total flux and the emission at the peak of the $EF_E$ spectrum (where $F_E$ is the differential energy flux) and beyond it. The reflection of this component is found to be weakly blurred relativistically and from a weakly ionized medium. The less luminous soft component dominates in soft X-rays, and its reflection is both more blurred and from a strongly ionized medium. In the fits, we determine the radial range of the soft reflection as between the inner disk truncation radius, $R_{\rm in}>R_{\rm{ISCO}}$, up to a transition radius, $R_{\rm tr}\equiv R_{\rm in}+\Delta R$. The hard reflection is from radii $>R_{\rm tr}$. The fit parameters $R_{\rm in}$ and $\Delta R$ are determined predominantly by the soft and hard reflection, respectively. Each of the two reflection zones is characterized by its ionization parameter, $\xi\equiv 4\pi F_{\rm irr}/n$, where $F_{\rm irr}$ is the irradiating flux and $n$ is the density of the reflecting part of the disk. 
 
\begin{figure}[t!]
\centerline{\includegraphics[width=5.4cm]{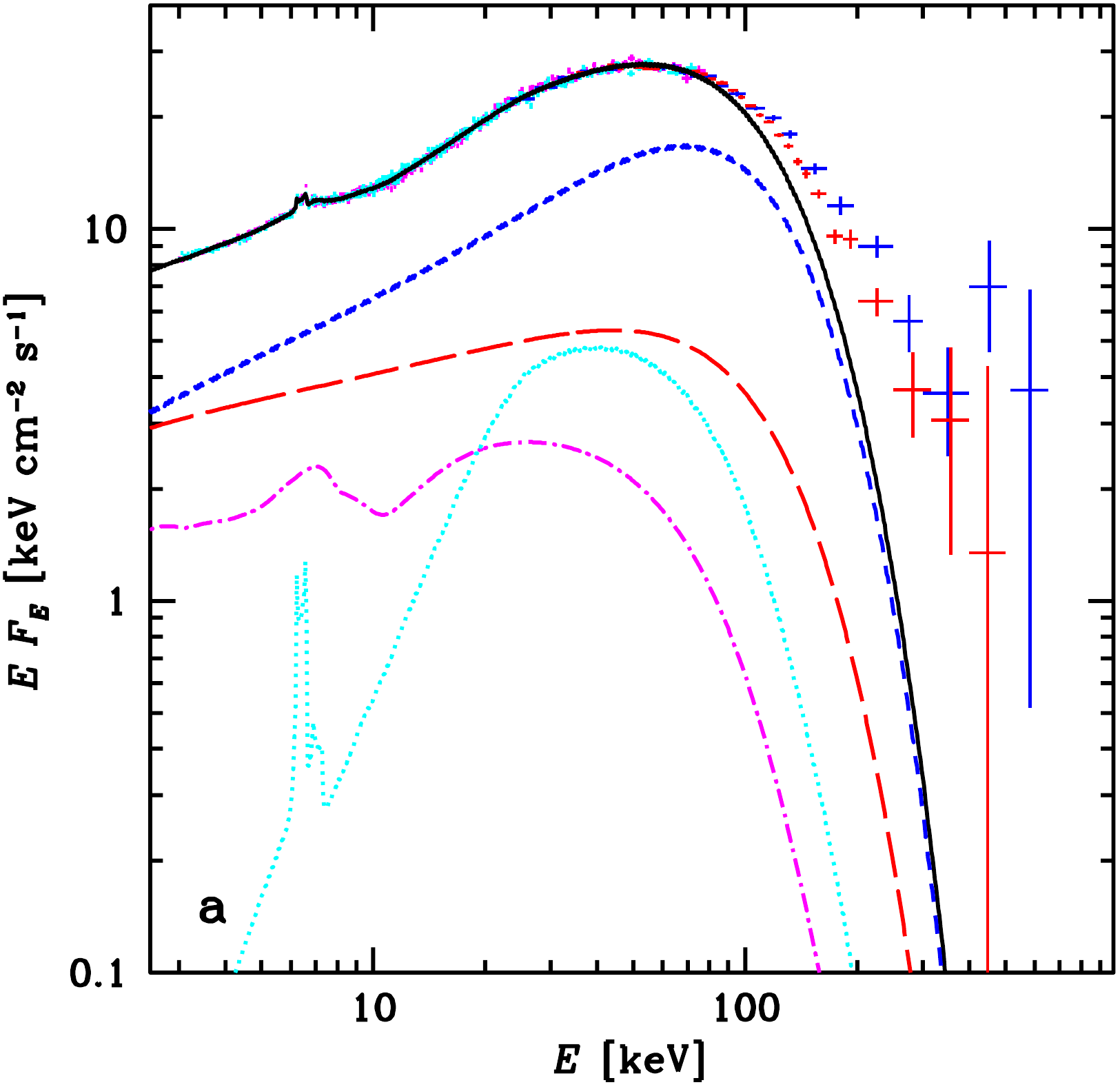}}
\centerline{\includegraphics[width=5.4cm]{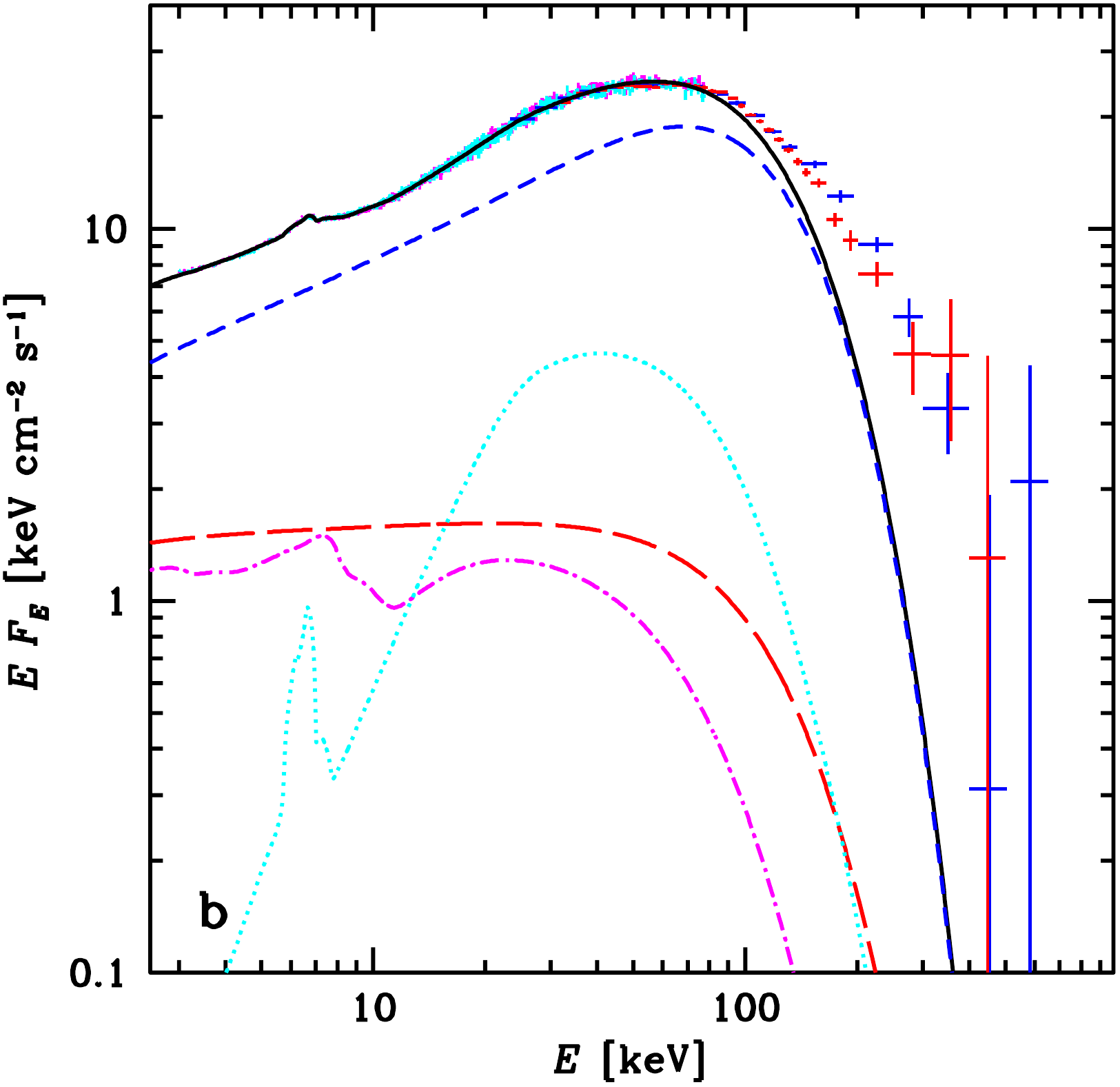}} 
  \caption{The \nustar (magenta and cyan), SPI (blue) and ISGRI (red) unfolded spectra of epochs (a) 1, (b) 2, fitted in the 3--80\,keV range with the two-component model with thermal electrons (black solid curves), but shown in the entire ranges of the data. The spectra are normalized to the \nustar A module. The fits significantly underpredict the spectra at $\gtrsim$100\,keV. The softer and harder Comptonization components are shown by  the red long dashes and blue short dashes, and the corresponding reflection components are shown by the magenta dot-dashes and cyan dots, respectively.
}\label{eeuf_th}
\end{figure}

However, the locations of the two Comptonizing plasmas (hard and soft) is not determined by the fits. A {\it possible\/} geometry is shown in fig.\ 4 of \zdz. There, the soft component forms a corona covering the disk between $R_{\rm in}$ and $R_{\rm tr}$, and its reflection is from the disk underneath. The hard component forms either a hot flow or a slow jet sheath with a large scale height at radii $<R_{\rm in}$. Its inner location is supported by its dominance of the total luminosity. Since the disk between $R_{\rm in}$ and $R_{\rm tr}$ is covered by the soft corona, the reflection of the hard Comptonization component is from remote parts of the disk, at $R>R_{\rm tr}$. While this geometry is not unique, our results below do not depend on it.

\renewcommand{\arraystretch}{0.9}
\begin{table*}
\caption{The results of spectral fitting for our two-component coronal model (soft thermal and hard hybrid)
}
   \centering\begin{tabular}{ccccc}
\hline
Component & Parameter & Epoch 1 & Epoch 2 & Average\\
\hline
ISM absorption & $N_{\rm H}$ $[10^{21}]$\,cm$^{-2}$ & \multicolumn{3}{c}{1.4f}\\
\hline
Joint constraints & $i$ $[\degr$] & $70^{+3}_{-3}$ & $66^{+1}_{-1}$ & $65^{+3}_{-1}$\\
& $Z_{\rm Fe}\, [\sun]$ & $1.5^{+0.4}_{-0.1}$ & $1.2^{+0.3}_{-0.1}$ & $1.1^{+0.1}_{-0.1}$\\
\hline
Thermal Comptonization  & $y_{\rm th}$ & $0.54^{+0.01}_{-0.01}$ & $0.53^{+0.01}_{-0.02}$ & $0.68^{+0.01}_{-0.01}$\\
and reflection  &$\Gamma_{\rm th}$ & $1.80^{+0.02}_{-0.03}$ & $1.87^{+0.02}_{-0.02}$ & $1.75^{+0.02}_{-0.01}$\\
& $kT_{\rm e,th}$ [keV] & $12^{+1}_{-1}$ & $12^{+1}_{-1}$ & $20^{+1}_{-1}$\\
&$R_{\rm in}\, [R_{\rm g}]$ & $22^{+5}_{-5}$ & $31^{+9}_{-5}$ & $105^{+7}_{-67}$\\
& ${\cal R}_{\rm th}$ & $0.63^{+0.02}_{-0.13}$ & $0.42^{+0.01}_{-0.01}$ & $0.37^{+0.04}_{-0.05}$\\
& $\log_{10} \xi_{\rm th}$ & $3.61^{+0.05}_{-0.02}$ & $3.51^{+0.02}_{-0.03}$ & $3.51^{+0.02}_{-0.02}$\\
& $N_{\rm th}$ & $2.76^{+0.10}_{-0.38}$ & $2.90^{+0.01}_{-0.04}$ & $3.31^{+0.05}_{-0.17}$\\
\hline 
Hybrid Comptonization &$y_{\rm h}$ & $0.95^{+0.05}_{-0.01}$ & $0.85^{+0.01}_{-0.01}$ & $1.04^{+0.01}_{-0.01}$\\
and reflection &$\Gamma_{\rm h}$  & $1.33^{+0.01}_{-0.03}$ & $1.43^{+0.01}_{-0.01}$ & $1.39^{+0.01}_{-0.01}$\\
& $kT_{\rm e,h}$ [keV] & $21^{+2}_{-1}$ & $21^{+1}_{-1}$ & $33^{+5}_{-1}$\\
&$\gamma_{\rm min}$ & $1.24^{+0.06}_{-0.07}$ & $1.24^{+0.01}_{-0.01}$ & $1.31^{+0.08}_{-0.09}$\\
&$p$ & $3.06^{+0.54}_{-0.52}$ & $2.98^{+0.15}_{-0.01}$ & $3.55^{+0.20}_{-0.08}$\\
&$\Delta R\, [R_{\rm g}]$ & $850^{+60}_{-620}$ & $300^{+240}_{-80}$ & $<140$\\ 
& ${\cal R}_{\rm h}$ & $0.38^{+0.03}_{-0.01}$ & $0.31^{+0.01}_{-0.05}$ & $0.90^{+0.15}_{-0.05}$\\
& $\log_{10} \xi_{\rm h}$ & $2.47^{+0.07}_{-0.06}$ & $0^{+1.72}$ & $0^{+1.70}$\\
& $N_{\rm h}$ & $1.28^{+0.23}_{-0.02}$ & $1.09^{+0.01}_{-0.02}$ & $0.51^{+0.01}_{-0.02}$\\
\hline
& $\chi_\nu^2$  & 954/792 & 1759/1307 & 2275/2011\\
\hline
\end{tabular}
\tablecomments{
For {\tt plabs*tbabs(reflkerr+reflkerr\_bb)}. $Z_{\rm Fe}$ is the Fe abundance in solar units, $R_{\rm tr}=R_{\rm in}+\Delta R$, $R_{\rm out}=10^3 R_{\rm g}$, $N_{\rm th,h}$ is the flux density of a Compton component @1\,keV, $y_{\rm th,h}$ is the Compton parameter (calculated accurately for spherical geometry), $\Gamma_{\rm th,h}$ is the power-law index fitted to a Compton component in the 2--10\,keV range (not a free parameter), and ${\cal R}_{\rm th,h}$ is the reflection fraction. }
\label{t_fits}
\end{table*}

\subsection{Fit results}
\label{results}

\begin{figure*}[t!]
\centerline{
\includegraphics[width=5.4cm]{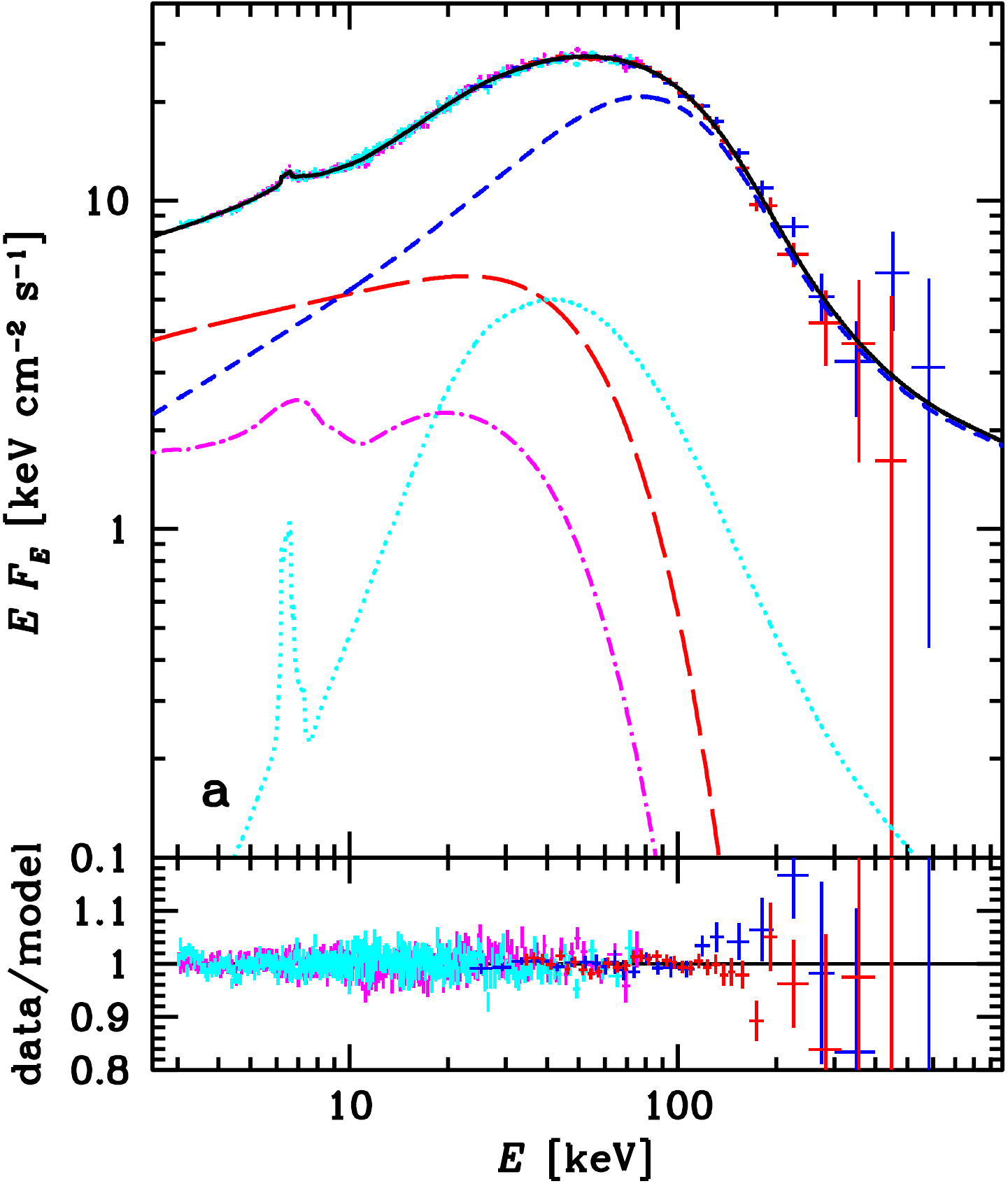}
\includegraphics[width=5.4cm]{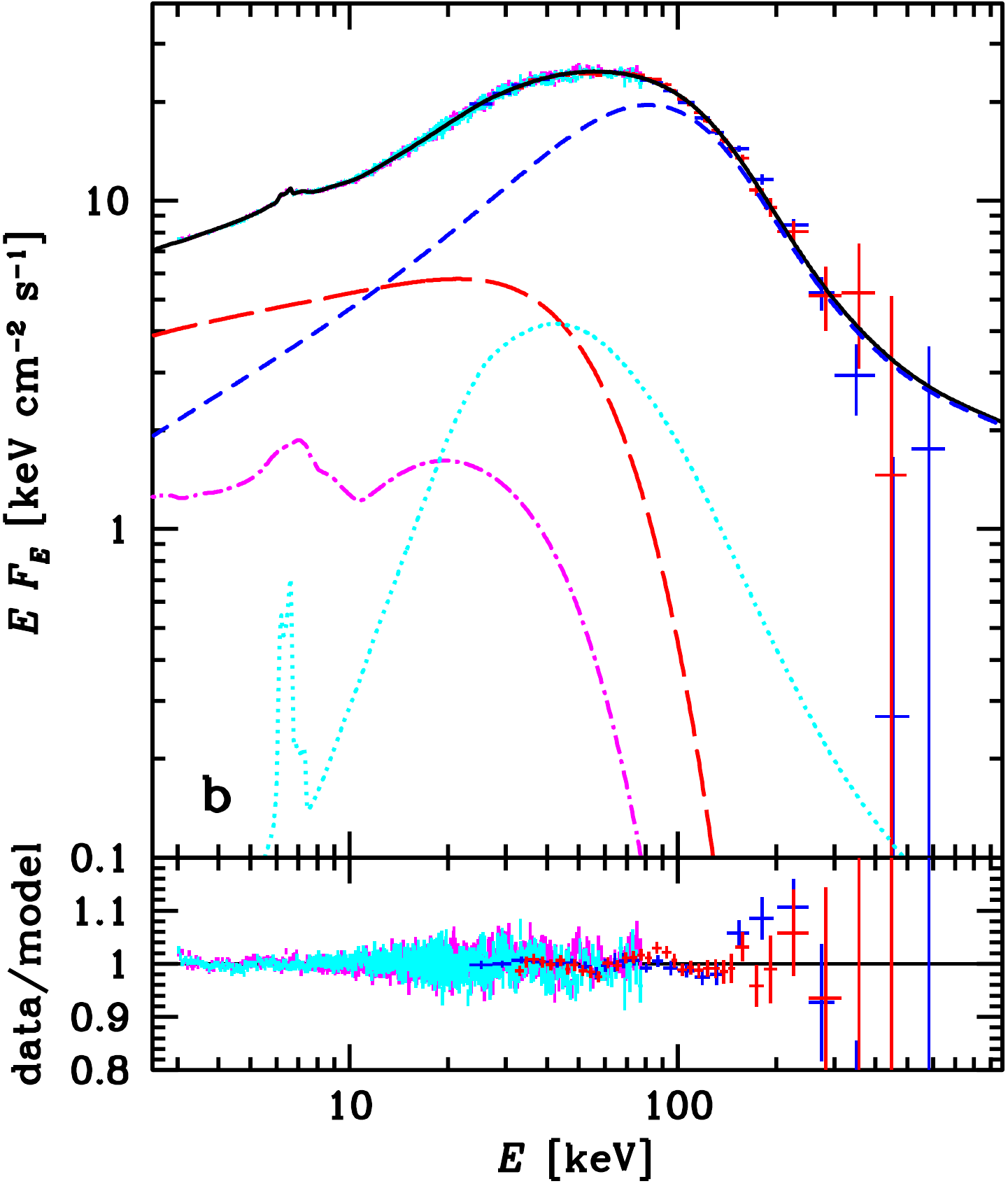} 
\includegraphics[width=6.1cm]{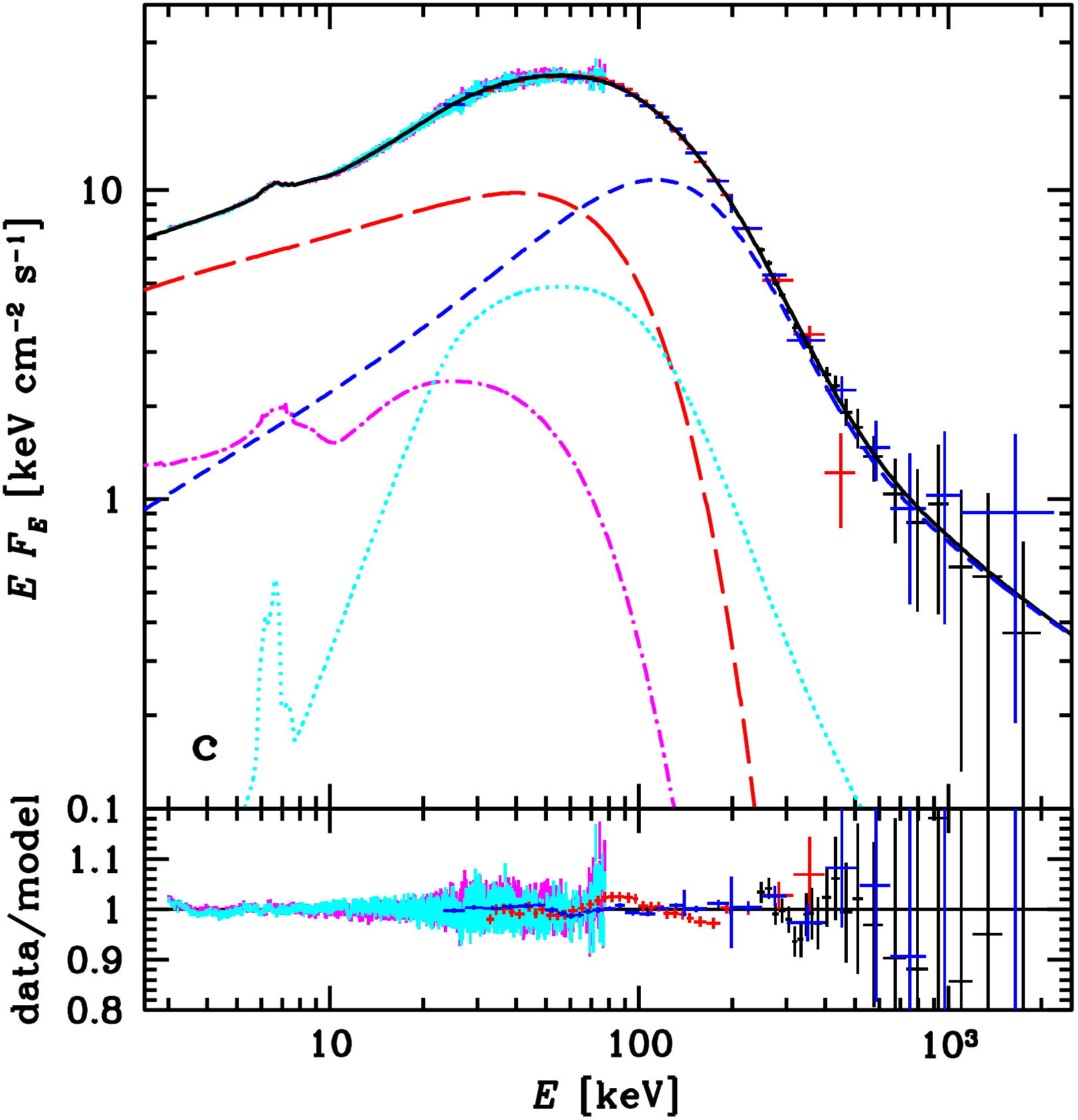}}
  \caption{The \nustar (magenta and cyan), SPI (blue), ISGRI (red) and PICSiT (black; for A only) unfolded spectra and data-to-model ratios of the epochs (a) 1, (b) 2, (c) A, fitted in the full ranges with the two-component coronal model (black solid curves), see Table \ref{t_fits}. The spectra are normalized to the \nustar A one. The softer (red long dashes) and harder (blue short dashes) Comptonization components are for thermal and hybrid electrons, respectively. The corresponding reflection components are shown by the magenta dot-dashes and cyan dots, respectively.
}\label{eeuf_ratio}
\end{figure*}

We first fit the \nustar data for epochs 1 and 2 assuming that the two Comptonization zones have the same electron temperature. We obtain results similar to those in table 2 in \zdz, with $R_{\rm in}\gg R_{\rm ISCO}$. The inclinations obtained from the fits are within $i\approx 60$--$70\degr$, which agrees with the binary and jet measurements (Section \ref{intro}). As in \zdz, we do not impose the same inclination for each data set, in order to demonstrate that each of them gives $i$ compatible with the observational constraints, as well as to account for possible flaring and precession of the disk. We then include the SPI and ISGRI data and fit them simultaneously with \nustar up to 80\,keV. We find an excellent agreement of those data with \nustar. In the next step, we include the data $>$80\,keV, and plot the resulting $EF_E$ spectra in Figure \ref{eeuf_th}. We find that both the SPI and ISGRI data at $E\gtrsim 100$\,keV are significantly above the model fitted at 3--80\,keV, while they are in a good mutual agreement. In particular, the spectra measured by the SPI and ISGRI extend up to 400\,keV with significant non-zero fluxes, while the models fitted in the 3--80\,keV range predict only very weak such emission.  

These results show that fitting the \nustar data with thermal Comptonization in order to determine the rate of e$^\pm$ pair production and constrain its equilibrium with the rate of pair annihilation in accreting BHs \citep{Fabian15} may be unreliable. At least in \source, the \nustar spectra fitted by models with a thermal electron distribution do not provide reliable predictions for the spectra around 511\,keV, and thus for the pair production rate. This conclusion agrees with that of \citet{Coppi99} and \citet{Fabian17}, who found that allowing for a hybrid electron distribution reduces the minimum electron temperature at which significant pair production can take place. 

We then fit the broad-band spectra for all three epochs with the thermal Comptonization model, but allowing $kT_{\rm e}$ of the two Comptonization clouds to be different. The resulting fits remain rather poor. In particular, we obtain $\chi^2_\nu\approx 984/794$, 1835/1309, 2437/2013 for epochs 1, 2, A, respectively, with significant residuals at the highest energies. These models also require very strong reflection from the harder component, e.g., ${\cal R}_{\rm h}\approx 2.8$, 3 (constrained to $\leq$3) for epochs 2 and A, respectively, which puts in question their physical reality. 

\begin{figure}
\centering
\includegraphics[width=6.8cm]{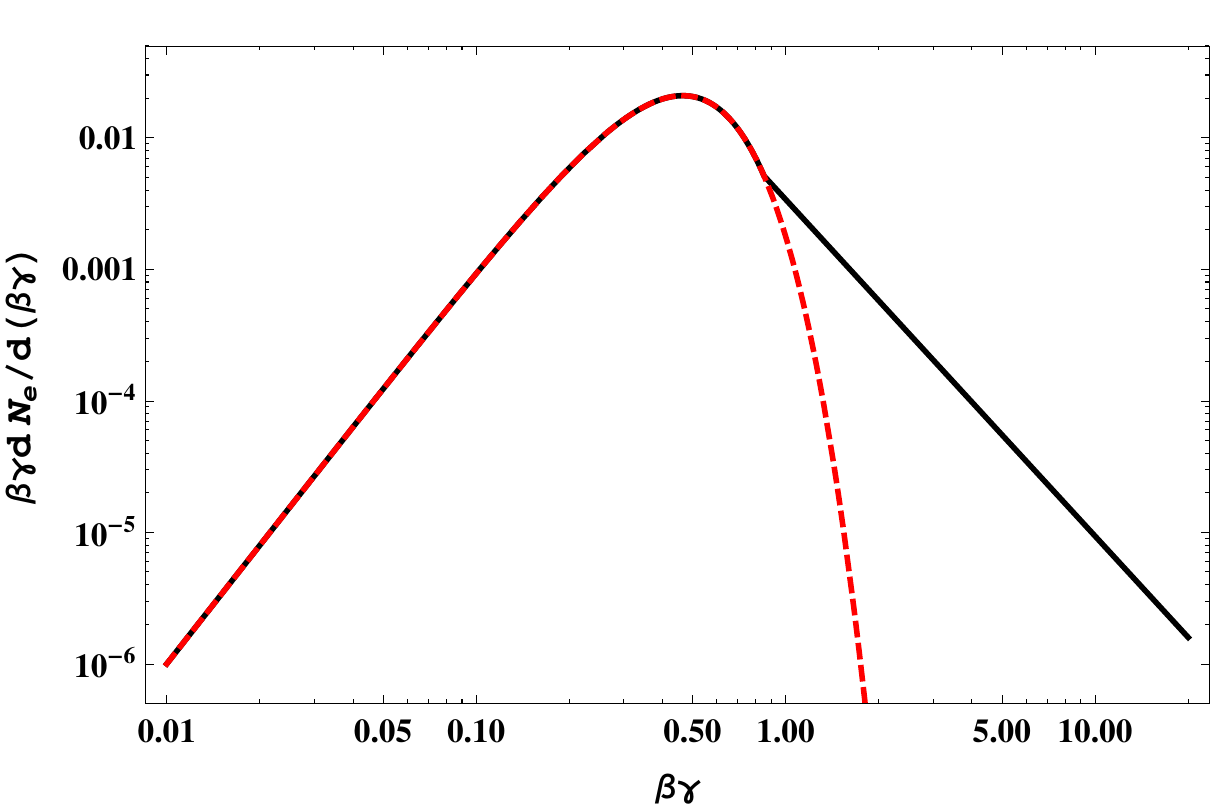}
  \caption{The electron distribution of hard Comptonization component of the best fit to epoch A, with $kT_{\rm e}=33$\,keV and a power-law with $p=3.55$ above $\beta\gamma$ corresponding to $\gamma_{\rm min}=1.31$. The dashed curve gives the corresponding pure Maxwellian distribution.
}\label{eldist}
\end{figure}

We then allow for the presence of a non-thermal tail in the electron distribution of the hard component, which adds two free parameters, $\gamma_{\rm min}$ and $p$. We find that this strongly improves the fits, with the spectra and data/model ratios shown in Figure \ref{eeuf_ratio} and the parameters given in Table \ref{t_fits}. With the new values of $\chi^2$, the probabilities for the improvement to be by chance estimated by the F-test for epochs 1, 2, A equal $\approx 5\times 10^{-6}$, $10^{-12}$, $10^{-30}$ respectively. We find $\gamma_{\rm min}\approx 1.2$--1.4 and $p\approx 2.5$--3.7. The index for the average spectrum is somewhat steeper than in the case of the individual observations. This may be due to the \integral and \nustar observations not covering exactly the same periods with different exposures and/or the high-energy tail weaker than in the case of epochs 1--2 during some of the other \integral observations (as observed; \citealt{Roques19}). Figure \ref{eldist} shows the electron distribution corresponding to the best fit to epoch A. The tail contains 8\% of the electrons and 40\% of their energy.

We have added a high-energy tail to the hard component only since its Comptonization dominates at high energies in modelling the spectra of epochs 1 and 2, which is also the case for all the observations studied in \zdz. Still, we have tested the effect of adding a tail also to the soft component, but found that while its presence is allowed (and likely), it leads to no decrease of $\chi^2$ for either epoch 1 or 2. For the average spectrum, we found a modest decrease of $\Delta\chi^2\approx -9$ with almost no change to the other fit parameters. The parameters of the tail to the soft component are poorly constrained, and the model becomes considerably more complicated. Thus, we have opted for not including that tail.

Apart from the presence of the non-thermal tail, the results shown in Table \ref{t_fits} for epochs 1 and 2 are very similar to those for the \nustar observations 2 and 4 analyzed by \zdz, with the Fe abundance close to solar and the disk being substantially truncated. On the other hand, the relatively large value of $R_{\rm in}$ for epoch A may be biased by the averaging, and we do not claim it corresponds to the actual average inner radius. 

Instead of adding a non-thermal tail to the electron distribution, \citet{Roques19} were able to fit the SPI data ($\gtrsim$20\,keV) alone by thermal Comptonization, reflection and an e-folded power law. However, the \nustar data (3--79\,keV) for this source require two reflection components \citep{Buisson19} as well as two incident continua with different shapes (\zdz), which we have assumed in our joint fitting. We can still add another spectral component to our two-component thermal-Compton fit described above. We have found that an added e-folded power law has a very low spectral index, $\Gamma\sim 0$. In the absence of self-absorption (which is ruled out at high energies), a spectrum as hard can be obtained only from thermal Comptonization close to saturation, with a pronounced Wien peak. We have thus replaced the e-folded power law by thermal Comptonization using the {\tt thcomp} model \citep{Z20_thcomp}, which is well suited for saturated Comptonization. For the average spectrum, we have obtained a reasonable fit with $\Gamma\approx 1.2$ and $kT_{\rm e}\approx 68$\,keV for that additional component. Still, this model yields $\chi^2_\nu\approx 2324/2010$, which is significantly higher than the model with a non-thermal tail (Table \ref{t_fits}), with $\Delta \chi^2\approx +49$, as well as the data are much above the model at $E\gtrsim 0.5$\,MeV. The two models are not nested, so the F-test cannot be used. Instead, the Akaike information criterion \citep{Akaike78} can be used, which yields the relative likelihood of the model with the additional component compared to that with the tail of $\approx\! 8\times 10^{-12}$, where we used it as given by eq.\ (1) in \citet{Dzielak19}. Thus, we see that an addition of a high-energy tail to the electron distribution yields a much more likely fit than that with an additional spectral component.

\section{Pair production}
\label{pair}

\begin{figure}
\centering
\includegraphics[width=6.cm]{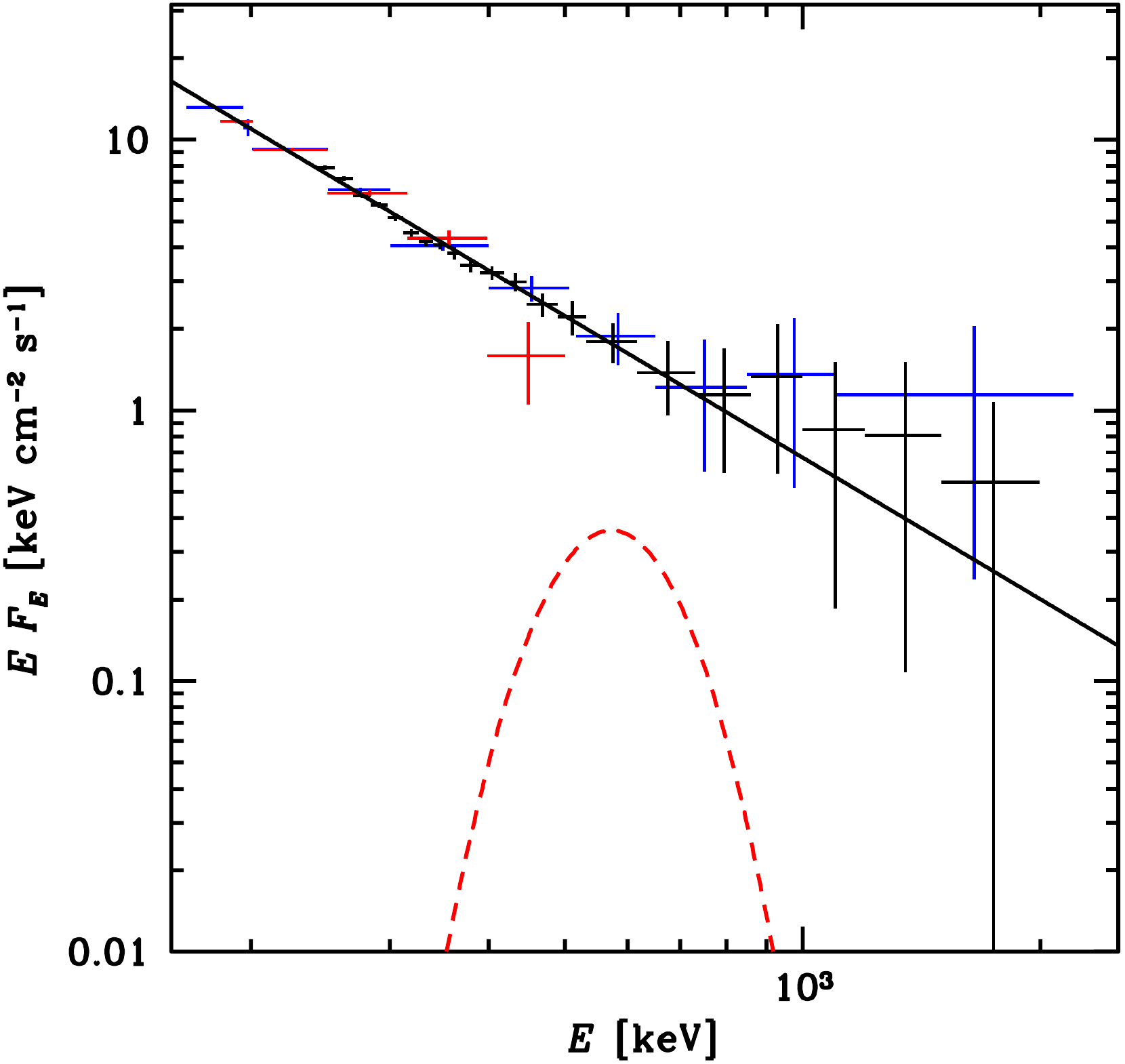} 
  \caption{The average data above 160\,keV (normalized to the SPI) fitted with a power law: SPI (blue), ISGRI (red), PICsIT (black). Including an e$^\pm$ pair annihilation line in the fit yields its null normalization. We plot the maximum line allowed by the data at $2\sigma$.
}\label{power_average}
\end{figure}

We express the photon-photon pair production rate through the differential photon density at its threshold, $m_{\rm e}c^2$, and express the photon energies as $\epsilon\equiv E/(m_{\rm e}c^2)$. We denote that density, ${\rm d}n/{\rm d}\epsilon$, at $\epsilon=1$ as $n_1$. In order to estimate it, we assume a simple model of a sphere with a radius, $R$, a uniform photon density, and the average photon escape time of $\approx \!(R/c) (1+g_{\rm C}\tau_{\rm T}/3)$, where $\tau_{\rm T}\equiv R n_{\rm e}\sigma_{\rm T}$, $g_{\rm C}\equiv \sigma_{\rm C}/\sigma_{\rm T}$, $\sigma_{\rm T}$ and $\sigma_{\rm C}$ are the Thomson and Compton cross sections, respectively, $n_{\rm e}$ is the total density of electrons and positrons, and $g_{\rm C}(\epsilon=1)\approx 0.43$. This yields
\begin{equation}
n_1\approx F_E(511{\rm keV})\frac{3(1+g_{\rm C}\tau_{\rm T}/3)d^2}{R^2 c},
\label{n1}
\end{equation}
where $F_E$ is in units of keV\,cm$^{-2}$\,s$^{-1}$\,keV$^{-1}$. Given that the spectra in the vicinity of 511\,keV have approximate power-law shapes, see Figures \ref{eeuf_ratio}, \ref{power_average}, we assume ${\rm d}n/{\rm d}\epsilon= n_1 \epsilon^{-\Gamma}$. Thus, the formalism developed in this section is independent of the hybrid model fitted to the broad-band spectra. We further approximate the photon field within the source as isotropic, allowing us to use the absorption coefficient for photon-photon pair production by \citet{GS67}\footnote{Corrected in \citet{Brown73}. See also the corrected first three coefficients of the series expansion in \citet{Zdziarski88}; the fourth is in error.}. The absorption coefficient for photons crossing a power-law isotropic photon field without cutoffs is given by eqs.\ (B3) and (B5) of \citet{Svensson87}, which corresponds to the radial optical depth of
\begin{equation}
\tau_{\gamma\gamma}(\epsilon)\approx \frac{7\epsilon^{\Gamma-1} n_1\sigma_{\rm T} R}{6\Gamma^{5/3} (1+\Gamma)}. 
\label{taugg}
\end{equation}
In the absence of cutoffs, the total pair production rate is infinite. Here, we integrate the absorption coefficient corresponding to Equation (\ref{taugg}) over a power law with a sharp high-energy cutoff at $\epsilon_{\rm c}\equiv E_{\rm c}/(m_{\rm e}c^2)>1$, and include a factor of 1/2 to account for double counting of pair-producing photons. Note that this neglects the effect of pair absorption, which we will consider later. This yields,
\begin{equation}
\dot n_{\gamma\gamma}\approx \frac{7 C \ln\epsilon_{\rm c}}
{6\Gamma^{5/3} (1+\Gamma)} n_1^2 \sigma_{\rm T} c.
\label{pprate}
\end{equation}
It depends logarithmically on the unknown high-energy cutoff. Equation (\ref{pprate}) also requires that the power law extends at low energies to at least $\epsilon=\epsilon_{\rm c}^{-1}$. The neglect of the cutoff in calculating the absorption coefficient means that the rate calculated as above is overestimated, which we account for by the factor $C<1$. At $\epsilon_{\rm c}=4$ and $\Gamma=3$, 4, $C\approx 0.71$, 0.78, respectively. Given the lack of an observed high-energy cutoff, the actual value of $\epsilon_{\rm c}$ remains unknown, but the average spectrum shown in Figure \ref{power_average} implies $\epsilon_{\rm c}\gtrsim 4$. 

We assume the produced pairs lose quickly their energy in excess of the thermal one in Compton and Coulomb scattering and thermalize. The annihilation rate by thermal e$^\pm$ is
\begin{equation}
(\dot n_+)_{\rm ann} =\frac{3}{32} u(2-u) n_{\rm e}^2 \sigma_{\rm T}c g_{\rm A}(\theta),
\label{annrate}
\end{equation}
where $u\equiv 2 n_+/n_{\rm e}$ is the pair abundance, $n_+$ is the positron density, $\theta\equiv kT_{\rm e}/m_{\rm e}c^2$ and $g_{\rm A}$ is the correction to the Born rate, which can be fitted as \citep{Svensson82a}
\begin{equation}
g_{\rm A}\approx \left[1+2\theta^2/\ln(2\eta_{\rm E}\theta +1.3)\right]^{-1},
\label{gA}
\end{equation}
where $\eta_{\rm E}\equiv \exp(-\gamma_{\rm E})\approx 0.5616$ and $\gamma_{\rm E}$ is Euler's constant. Here, we neglect the small effect of annihilation of e$^\pm$ from the non-thermal tail of the fitted distribution, whose tail contains only a small fraction of all e$^\pm$.

We then consider escape of pairs, which is very likely to be present as advection to both the BH and the jet/wind. We parametrize it by the ratio of the light travel time across the source to the time scale of pair escape, $\beta_{\rm esc}\equiv (R/c)/[n_+/(\dot n_+)_{\rm esc}]<1$ (following eq.\ 18 of \citealt{Zdziarski85}). This gives
\begin{equation}
(\dot n_+)_{\rm esc}=\frac{1}{2\tau_{\rm T}}\beta_{\rm esc} u n_{\rm e}^2\sigma_{\rm T}c.
\label{escape}
\end{equation}
The rates of annihilation and escape become equal for
\begin{equation}
\beta_{\rm esc}=\frac{3}{16}(2-u)g_{\rm A}(\theta) \tau_{\rm T}
\label{beta_eq}
\end{equation}
(not assumed below). Pair equilibrium corresponds to $\dot n_{\gamma\gamma}= (\dot n_+)_{\rm ann}+(\dot n_+)_{\rm esc}$. (This neglects the photon-particle and particle-particle pair production, which, as we have checked, are not important for this source.) Its solution gives the radius of
\begin{equation}
R_{\rm eq}\approx \frac{4(7 C \ln \epsilon_{\rm c})^{\frac{1}{2}}\sigma_{\rm T} d^2 (1+0.14\tau_{\rm T}) F_E(511{\rm keV})}{\Gamma^{5/6}[u(\Gamma+1)]^{\frac{1}{2}}\tau_{\rm T}c \left[(2-u)g_{\rm A}+\frac{16\beta_{\rm esc}}{3\tau_{\rm T}}\right]^{\frac{1}{2}}},
\label{radius}
\end{equation}
which is $\appropto u^{-1/2}$. The minimum radius at which pair equilibrium can be established, which we denote as $R_{\rm pair}$, corresponds to the pair dominance, i.e., $u\rightarrow 1$. Also, for known $\tau_{\rm T}$ and radius, e.g., setting $R_{\rm eq}=R_{\rm in}$ of Table \ref{t_fits}, we can solve for $u$. For a given $u$, the optical depth to pair production can be calculated using Equation (\ref{taugg}),
\begin{equation}
\tau_{\gamma\gamma}(\epsilon)\approx \frac{\epsilon^{\Gamma-1}\tau_{\rm T} (7 u)^{\frac{1}{2}} \left[(2-u)g_{\rm A}+\frac{16\beta_{\rm esc}}{3\tau_{\rm T}}\right]^{\frac{1}{2}} } {8\Gamma^{5/6}(\Gamma+1)^{\frac{1}{2}}(C \ln \epsilon_{\rm c})^{\frac{1}{2}} },
\label{taugg1}
\end{equation}
which is maximized at $u=1$ (at which $R=R_{\rm pair}$).

\renewcommand{\arraystretch}{0.9}
\setlength{\tabcolsep}{2pt}
\begin{table}\centering
\caption{High-energy and e$^\pm$ pair parameters 
}
\vskip -0.4cm                               
\begin{tabular}{lcccccc}
\hline
Epoch &$\Gamma$ &$\frac{E F_E(511{\rm keV})}{{\rm keV\,cm}^{-2}\,{\rm s}^{-1}}$ &$\tau_{\rm T}$ &$\frac{R_{\rm pair}}{R_{\rm g}}$ &$\left.\tau_{\gamma\gamma}\right(u\!=\!1)$ &$u(R_{\rm in})$\\
\hline
1 & $3.60^{+0.26}_{-0.24}$ & $2.48_{-0.54}^{+0.62}$ &5.7 & 3.8 & 0.34& 0.017\\
2 & $3.70^{+0.22}_{-0.20}$ & $2.35_{-0.43}^{+0.48}$ &5.1 & 3.7 & 0.29& 0.008\\
A & $3.74^{+0.05}_{-0.04}$ & $2.14_{-0.09}^{+0.09}$ &4.0 & 3.7 & 0.23& 0.0007\\
\hline
\end{tabular}
\tablecomments{
$\Gamma$ and $E F_E(511{\rm keV})$ are for the fits at $E\gtrsim$160\,keV. The Thomson depth of the hard component, $\tau_{\rm T}$, the radius for $u=1$, $R_{\rm pair}$, the photon-photon optical depth at $u=\epsilon=1$, $\tau_{\gamma\gamma}$, and $u$, the pair abundance for $R_{\rm eq}=R_{\rm in}$, are given for the best-fit parameters, and $M=8\msun$, $\epsilon_{\rm c}=4$, $\beta_{\rm esc}=0.3$, $C=0.75$.}
\label{pairs}
\end{table}

We then calculate the total observed flux of photons from pair annihilation as $\dot N_{\rm ann}= 2(\dot n_+)_{\rm ann}(4\pi R^3/3)/[4\pi d^2(1+0.14\tau_{\rm T})]$ (in units of cm$^{-2}$\,s$^{-1}$), where the factor of 2 accounts for two photons produced in each annihilation event and the flux is reduced by scattering (and we neglect secondary pair production by the annihilation photons). Setting $R=R_{\rm eq}$, we find
\begin{equation}
\dot N_{\rm ann}\approx \frac{(7 C u\ln \epsilon_{\rm c})^{\frac{1}{2}}(2-u)\tau_{\rm T} F_E(511{\rm keV})}{4\Gamma^{5/6} (\Gamma+1)^{\frac{1}{2}}\left[(2-u)g_{\rm A}+\frac{16\beta_{\rm esc}}{3\tau_{\rm T}}\right]^{\frac{1}{2}}}.
\label{obs_ann}
\end{equation}
The above rate assumes that annihilation of pairs escaping the hot plasma is negligible. In the limit of $kT_{\rm e}\ll m_{\rm e}c^2$ (which is satisfied in \source), the observed energy spectrum ($F_{E,{\rm ann}}$ in units of keV\,cm$^{-2}$\,s$^{-1}$\,keV$^{-1}$) from pair annihilation becomes \citep{SLP96}
\begin{equation}
F_{E,{\rm ann}}\approx \dot N_{\rm ann}\frac{\epsilon^{5/2}}{(\pi \theta)^{1/2}}\exp\left[-\frac{(\epsilon-1)^2}{\epsilon\theta}\right].
\label{anspec}
\end{equation}
Here, we assumed that the annihilating e$^\pm$ are purely thermal. This is approximately satisfied for this source; e.g., the steady-state electron distribution for epoch A contains only 8\% of non-thermal e$^\pm$.

We now use the data from \source to determine the values of the parameters related to pair production. We have fitted the spectra of the three epochs at $E\gtrsim 160$\,keV by power laws, and show the results in Table \ref{pairs} and in Figure \ref{power_average} for the average spectrum. Then we use equation (\ref{radius}) assuming $\epsilon_{\rm c}=4$, $C=0.75$ and $\beta_{\rm esc}=0.3$, and use the values of $\tau_{\rm T}$ obtained from the broad-band fits. The obtained radii for the pair dominance are given in Table \ref{pairs}, expressed in units of the gravitational radius for the assumed BH mass of $8\msun$. We find $R_{\rm pair}\approx 4 R_{\rm g}$ and, from equation (\ref{taugg1}), the corresponding $\tau_{\gamma\gamma} (\epsilon=1) \approx 0.3$. At such low radii, GR effects would increase the photon energies and density, and lead to some increase of $R_{\rm pair}$. Also, a possible flattening of the photon distribution at highest energies, as fitted in Figure \ref{eeuf_ratio}c, would also increase $R_{\rm pair}$ by a modest factor. On the other hand, attenuation of the power law at $\epsilon\gtrsim 1$, if present, would decrease $R_{\rm pair}$. 

The actual radii of the plasmas emitting in the vicinity of 511\,keV remain unknown. Adopting the hybrid Comptonization model, this emission is produced by the hard spectral component, which reflection is only weakly relativistically blurred, see Section \ref{fits}. As discussed there, this argues for its size being relatively large, $\sim\! R_{\rm in}$ or more. The values of $R_{\rm in}$ are larger by an order of magnitude than $R_{\rm pair}$. By solving $R_{\rm eq}=R_{\rm in}$, we find $u\ll 1$, see Table \ref{pairs}. Thus, the hot plasma within $R_{\rm in}$ appears not to be dominated by pairs. In this case, $\tau_{\gamma\gamma}\ll 1$, and our neglect of pair absorption is justified. The low pair density in this source is also corroborated by the lack of any apparent cutoff above 511\,keV in the observed spectrum.  

A low pair density within the emitting plasma is further strongly supported by results of a fit to the average spectrum including an annihilation feature. We have found that the best-fit normalization of the feature is null. For that, we assumed the fitted value of $kT_{\rm e,h}=33$\,keV; however, the null best fit value is obtained at any $kT_{\rm e}\lesssim 150$\,keV. We show the annihilation spectrum corresponding to the $2\sigma$ upper limit in Figure \ref{power_average}. This limit corresponds to $\lesssim$0.18 of the normalization corresponding to the pair-dominated plasma ($u=1$), which, in turn, corresponds to $u\lesssim 0.014$ and $R_{\rm eq}\gtrsim 24 R_{\rm g}$, in agreement with the constraints from the fits to epochs 1 and 2 assuming $R\sim R_{\rm in}$.

\section{Discussion}
\label{discussion}

In the hard state, we have obtained the disk inner truncation radii of $R_{\rm in}\approx 20$--$30 R_{\rm g}$ (from spectral fitting to the individual observations), while the equilibrium radii corresponding of pair dominance assuming the fitted Thomson optical depths are $R_{\rm pair}\sim 4 R_{\rm g}$. If the size of the plasma emitting around 511\,keV is comparable to $R_{\rm in}$ (or larger), as we argued above, its pair abundance is low. Still, the obtained values of $R_{\rm pair}$ hint at the importance of pair production in regulating the physical state of the flow. If pairs were just unimportant, the minimum radius limited by pair equilibrium could have any value, but instead we find $R_{\rm pair}\sim R_{\rm ISCO}$, suggesting that pair production somehow knows the fundamental size scale of the system. The pairs produced by the non-thermal photon tail may contribute to regulation of the temperature of the thermal component of the electron distribution \citep{Coppi99,Fabian17}.

Alternatively, the obtained low values of the electron temperature, $kT_{\rm e}\sim 20$--30\,keV, could be due to the energy balance in the flow itself.  \citet{PV09} and \citet{MB09} suggested $T_{\rm e}$ can be regulated by thermalization in a synchrotron boiler \citep{GGS88}. Such regulation also takes place in advection-dominated accretion flows (ADAFs), but the electron temperatures of such flows are much higher, e.g., \citet{YN14}. On the other hand, the existing ADAF models consider only thermal electrons, while the presence of non-thermal tails can greatly increase the synchrotron cooling (see below), which effect, could, in principle, reconcile this model with the data. Some studies of extended accretion flows with hybrid electrons were performed by \citet{VPV13}, but without including heating from ions.

An important diagnostic, strongly confirming the low pair dominance independently of the pair equilibrium calculations, is the absence of an annihilation feature in the observed spectrum. Purely thermal plasmas do not show distinct pair annihilation lines even if dominated by pairs \citep{Z86,Stern95,Svensson96}. However, the data for this source rule out purely thermal plasmas, see Figure \ref{eeuf_th}. On the other hand, hybrid plasmas can emit X-ray spectra well reproduced by Comptonization on mostly thermal electrons, but still exhibit relatively narrow annihilation features. The absence of such a feature in the average spectrum therefore implies the pair abundance to be very low.

The pair equilibrium formalism developed in Section \ref{pairs} is independent of the specific hybrid model fitted to the data. In particular, the pair production rate is calculated for a general photon power-law spectrum, and is then independent of the thermal-Compton parts of the observed spectra. Thus, this rate remains approximately valid even if the high-energy tail is produced in a spatially different region (e.g., some region of the jet) than that of the thermal Comptonization, and it depends mostly on the characteristic size of that region. However, the rates of pair annihilation and escape strongly depend on the Thomson optical depth of the pair-producing plasma, which then yields $R_{\rm pair} \appropto \tau_{\rm T}^{-1}$. We used the values of $\tau_{\rm T}$ of the hard Comptonization region from our hybrid fits, which then yields $R_{\rm pair}$ as given in Table \ref{pairs}.  Alternative models with lower $\tau_{\rm T}$ of the pair-producing region would then yield larger $R_{\rm pair}$. Then, the pair abundance would depend on both that $R_{\rm pair}$ and the size of that region. While we have been unable to find acceptable models to the data with a separate region emitting the photon high-energy tail, we cannot rule out their existence.

The best-fit steady-state power-law indices of the electron tail are in the range of $p\approx 3.0$--3.6. The Compton (in the Thomson limit) and synchrotron energy losses steepen the distribution of accelerated relativistic electrons by unity, giving their index as $p_{\rm acc}=p-1$ (though we caution that some of the emitting electrons in our case are only mildly relativistic). Thus, the implied values of $p_{\rm acc}$ are in good agreement with those from shock acceleration. In particular, indices $p_{\rm acc}\approx 2.2$--2.4 are found for collisionless shocks \citep{Sironi15}. On the other hand, magnetic reconnection can readily give particle spectra of similar indices \citep{Sironi16, Ball18, Sironi20}. 

The presence of high-energy tails in the electron distribution likely results in the synchrotron emission of the electrons being greatly enhanced with respect to the Maxwellian case \citep{WZ01,VVP11,VPV13, PV14}. This is likely to result in the synchrotron emission becoming an important source of the seed photons for Comptonization, with a typical self-absorption turnover at $\lesssim$10\,eV. We have thus run spectral models in which the seed photon temperature was set to 10\,eV, and found similar values of $\chi^2$ and minor differences in the best-fit parameters with respect to our main calculations (with $kT_{\rm bb}=200$\,eV). Thus, either the disk blackbody and synchrotron emission can provide the dominant seed photons, with the spectral fits at $>$3\,keV not allowing us to distinguish between these possibilities.

We note that the \nustar data were fitted by \citet{Buisson19} with $R_{\rm in}\approx 5 R_{\rm g}$ and $i\approx 30$--$35\degr$, both much lower than our values. As discussed in \zdz, the \nustar data for this source yield two separate solutions, one low-$i$, low-$R_{\rm in}$, and the other high-$i$, high-$R_{\rm in}$. However, the low-$R_{\rm in}$ one is ruled out by the binary and jet measurements, showing $i\geq 60\degr$ \citep{Torres20,Atri20}, leaving the high-$R_{\rm in}$ solution as applicable to this source.

\section{Conclusions}

We have studied two joint spectra from the 2018 hard-state of \source from \nustar and \integral. We have found an excellent agreement between the \nustar and SPI spectra in the overlapping region of 23--79\,keV, and an overall agreement between \nustar and ISGRI. We have found the resulting 3--650\,keV spectra are well fitted by a two-component Comptonization model, with each component having different spectral index and separate reflection regions. However, the harder component, dominant at high energies, is required to be emitted by hybrid electrons, whose distribution consists of a Maxwellian and a high-energy tail. The presence of the tail is required at a very high significance. A similar tail can be present also in the soft component; however, its parameters are poorly constrained and its presence would only weakly affect the fitting results due to the very low contribution of that component at high energies.

Except for the presence of the tail, the parameters fitted to the spectra are similar to those fitted by \zdz to the \nustar data of the observations taken within the same state of the source (in particular those of their epochs 2 and 4). In particular, we find the disk is truncated at $\approx$20--$30 R_{\rm g}$. The electron temperatures are low, $\approx$20--30\,keV for the harder component, while the optical depths are relatively large, $\tau_{\rm T}\approx 4$--5. 

We have also obtained an average hard-state spectrum of \source, based on the data spanning 53 d. This spectrum is quite similar to those of the two individual observations, but extends to $\approx$2\,MeV. It also includes the PICSiT data, which are found to be in an excellent agreement with those of the SPI.   

The average spectrum crosses 511\,keV, the threshold for e$^\pm$ pair production. We have developed a formalism calculating the rates of pair production, annihilation and escape in hybrid plasmas, and obtained formulae for the equilibrium radius at a given pair abundance and the optical depth to pair absorption. We have found that the minimum possible radius from the equilibrium condition, for a pair-dominated plasma, is $\approx\! 4 R_{\rm g}$. This is, however, much less than our estimates of the size of the hot plasma, $R\sim R_{\rm in}\sim 20$--$30 R_{\rm g}$. At such radii, the equilibrium pair abundance is $\ll$1. This conclusion is confirmed by the absence of an annihilation feature in the average spectrum. Even at its $2\sigma$ upper limit, the implied size of the hot plasma is $\gtrsim 20 R_{\rm g}$, corresponding to a low pair abundance. Also, we have found no steepening of the spectrum above 511\,keV, which would occur due to pair absorption if the plasma were pair dominated.

We consider it possible that pair production still acts as a thermostat, limiting the electron temperature. On the other hand, the low temperature can be due to an efficient cooling by synchrotron photons, which are copiously emitted by hybrid plasmas.

\section*{Acknowledgments}
We thank the referee for valuable comments. This work is based on observations with \integral, an ESA project with instruments and Science Data Center funded by ESA member states (especially the PI countries: Denmark, France, Germany, Italy, Switzerland, Spain; but also with a significant contribution from Poland) and with the participation of Russia and the USA. The \integral SPI project has been completed under the responsibility and leadership of CNES. The SPI team is grateful to ASI, CEA, CNES, DLR, ESA, INTA, NASA, and OSTC for their support. We acknowledge support from the International Space Science Institute (Bern), and the Polish National Science Centre under the grants 2015/18/A/ST9/00746, 2019/35/B/ST9/03944 and 2014/13/B/ST9/00570. AV and JP acknowledge the Academy of Finland grants 309308 and 333112, respectively.

\software{{\sc HEAsoft} (v6.25; HEASARC 2014), {\tt NuSTARDAS} (v.1.8.0), {\tt reflkerr} \citep{Niedzwiecki19}, {\tt xillverCp} (v.\ 1.4.3; \citealt{GK10,Garcia18}), {\tt ireflect} \citep{MZ95}, {\sc osa} (v.\ 11.1; \citealt{Courvoisier03}), {\sc xspec} \citep{Arnaud96}, {\tt compps} \citep{PS96}}.

\bibliography{allbib}{}

\begin{thebibliography}{}
\expandafter\ifx\csname natexlab\endcsname\relax\def\natexlab#1{#1}\fi
\providecommand{\url}[1]{\href{#1}{#1}}
\providecommand{\dodoi}[1]{doi:~\href{http://doi.org/#1}{\nolinkurl{#1}}}
\providecommand{\doeprint}[1]{\href{http://ascl.net/#1}{\nolinkurl{http://ascl.net/#1}}}
\providecommand{\doarXiv}[1]{\href{https://arxiv.org/abs/#1}{\nolinkurl{https://arxiv.org/abs/#1}}}

\bibitem[{{Akaike}(1978)}]{Akaike78}
{Akaike}, J.~E. 1978, Ann.\ Inst.\ Stat.\ Math., 30, 9,
  \dodoi{10.1007/BF02480194}

\bibitem[{{Anders} \& {Grevesse}(1989)}]{AG89}
{Anders}, E., \& {Grevesse}, N. 1989, \gca, 53, 197,
  \dodoi{10.1016/0016-7037(89)90286-X}

\bibitem[{{Arnaud}(1996)}]{Arnaud96}
{Arnaud}, K.~A. 1996, Astronomical Society of the Pacific Conference Series,
  Vol. 101, {XSPEC: The First Ten Years}, ed. G.~H. {Jacoby} \& J.~{Barnes}, 17

\bibitem[{{Atri} {et~al.}(2020){Atri}, {Miller-Jones}, {Bahramian}, {Plotkin},
  {Deller}, {Jonker}, {Maccarone}, {Sivakoff}, {Soria}, {Altamirano},
  {Belloni}, {Fender}, {Koerding}, {Maitra}, {Markoff}, {Migliari}, {Russell},
  {Russell}, {Sarazin}, {Tetarenko}, \& {Tudose}}]{Atri20}
{Atri}, P., {Miller-Jones}, J.~C.~A., {Bahramian}, A., {et~al.} 2020, \mnras,
  493, L81, \dodoi{10.1093/mnrasl/slaa010}

\bibitem[{{Ball} {et~al.}(2018){Ball}, {Sironi}, \& {{\"O}zel}}]{Ball18}
{Ball}, D., {Sironi}, L., \& {{\"O}zel}, F. 2018, \apj, 862, 80,
  \dodoi{10.3847/1538-4357/aac820}

\bibitem[{{Beloborodov}(2017)}]{Beloborodov17}
{Beloborodov}, A.~M. 2017, \apj, 850, 141, \dodoi{10.3847/1538-4357/aa8f4f}

\bibitem[{{Brown} {et~al.}(1973){Brown}, {Mikaelian}, \& {Gould}}]{Brown73}
{Brown}, R.~W., {Mikaelian}, K.~O., \& {Gould}, R.~J. 1973, \aplett, 14, 203

\bibitem[{{Buisson} {et~al.}(2019){Buisson}, {Fabian}, {Barret}, {F{\"u}rst},
  {Gandhi}, {Garc{\'\i}a}, {Kara}, {Madsen}, {Miller}, {Parker}, {Shaw},
  {Tomsick}, \& {Walton}}]{Buisson19}
{Buisson}, D.~J.~K., {Fabian}, A.~C., {Barret}, D., {et~al.} 2019, \mnras, 490,
  1350, \dodoi{10.1093/mnras/stz2681}

\bibitem[{{Cangemi} {et~al.}(2021){Cangemi}, {Beuchert}, {Siegert},
  {Rodriguez}, {Grinberg}, {Belmont}, {Gouiff{\`e}s}, {Kreykenbohm}, {Laurent},
  {Pottschmidt}, \& {Wilms}}]{Cangemi21}
{Cangemi}, F., {Beuchert}, T., {Siegert}, T., {et~al.} 2021, \aap.
\newblock \doarXiv{2102.04773}

\bibitem[{{Coppi}(1999)}]{Coppi99}
{Coppi}, P.~S. 1999, in Astronomical Society of the Pacific Conference Series,
  Vol. 161, High Energy Processes in Accreting Black Holes, ed. J.~{Poutanen}
  \& R.~{Svensson}, 375--403

\bibitem[{{Courvoisier} {et~al.}(2003){Courvoisier}, {Walter}, {Beckmann},
  {Dean}, {Dubath}, {Hudec}, {Kretschmar}, {Mereghetti}, {Montmerle},
  {Mowlavi}, {Paltani}, {Preite Martinez}, {Produit}, {Staubert}, {Strong},
  {Swings}, {Westergaard}, {White}, {Winkler}, \& {Zdziarski}}]{Courvoisier03}
{Courvoisier}, T.~J.~L., {Walter}, R., {Beckmann}, V., {et~al.} 2003, \aap,
  411, L53, \dodoi{10.1051/0004-6361:20031172}

\bibitem[{{Done} {et~al.}(2007){Done}, {Gierli{\'n}ski}, \& {Kubota}}]{DGK07}
{Done}, C., {Gierli{\'n}ski}, M., \& {Kubota}, A. 2007, \aapr, 15, 1,
  \dodoi{10.1007/s00159-007-0006-1}

\bibitem[{{Dzie{\l}ak} {et~al.}(2019){Dzie{\l}ak}, {Zdziarski}, {Szanecki}, {De
  Marco}, {Nied{\'z}wiecki}, \& {Markowitz}}]{Dzielak19}
{Dzie{\l}ak}, M.~A., {Zdziarski}, A.~A., {Szanecki}, M., {et~al.} 2019, \mnras,
  485, 3845, \dodoi{10.1093/mnras/stz668}

\bibitem[{{Fabian} {et~al.}(2017){Fabian}, {Lohfink}, {Belmont}, {Malzac}, \&
  {Coppi}}]{Fabian17}
{Fabian}, A.~C., {Lohfink}, A., {Belmont}, R., {Malzac}, J., \& {Coppi}, P.
  2017, \mnras, 467, 2566, \dodoi{10.1093/mnras/stx221}

\bibitem[{{Fabian} {et~al.}(2015){Fabian}, {Lohfink}, {Kara}, {Parker},
  {Vasudevan}, \& {Reynolds}}]{Fabian15}
{Fabian}, A.~C., {Lohfink}, A., {Kara}, E., {et~al.} 2015, \mnras, 451, 4375,
  \dodoi{10.1093/mnras/stv1218}

\bibitem[{{Garc{\'{\i}}a} \& {Kallman}(2010)}]{GK10}
{Garc{\'{\i}}a}, J., \& {Kallman}, T.~R. 2010, \apj, 718, 695,
  \dodoi{10.1088/0004-637X/718/2/695}

\bibitem[{{Garc{\'{\i}}a} {et~al.}(2018){Garc{\'{\i}}a}, {Steiner}, {Grinberg},
  {Dauser}, {Connors}, {McClintock}, {Remillard}, {Wilms}, {Harrison}, \&
  {Tomsick}}]{Garcia18}
{Garc{\'{\i}}a}, J.~A., {Steiner}, J.~F., {Grinberg}, V., {et~al.} 2018, \apj,
  864, 25, \dodoi{10.3847/1538-4357/aad231}

\bibitem[{{Gendreau} {et~al.}(2016){Gendreau}, {Arzoumanian}, {Adkins},
  {Albert}, {Anders}, {Aylward}, {Baker}, {Balsamo}, {Bamford}, {Benegalrao},
  {Berry}, {Bhalwani}, {Black}, {Blaurock}, {Bronke}, {Brown}, {Budinoff},
  {Cantwell}, {Cazeau}, {Chen}, {Clement}, {Colangelo}, {Coleman},
  {Coopersmith}, {Dehaven}, {Doty}, {Egan}, {Enoto}, {Fan}, {Ferro}, {Foster},
  {Galassi}, {Gallo}, {Green}, {Grosh}, {Ha}, {Hasouneh}, {Heefner}, {Hestnes},
  {Hoge}, {Jacobs}, {J{\o}rgensen}, {Kaiser}, {Kellogg}, {Kenyon}, {Koenecke},
  {Kozon}, {LaMarr}, {Lambertson}, {Larson}, {Lentine}, {Lewis}, {Lilly},
  {Liu}, {Malonis}, {Manthripragada}, {Markwardt}, {Matonak}, {Mcginnis},
  {Miller}, {Mitchell}, {Mitchell}, {Mohammed}, {Monroe}, {Montt de Garcia},
  {Mul{\'e}}, {Nagao}, {Ngo}, {Norris}, {Norwood}, {Novotka}, {Okajima},
  {Olsen}, {Onyeachu}, {Orosco}, {Peterson}, {Pevear}, {Pham}, {Pollard},
  {Pope}, {Powers}, {Powers}, {Price}, {Prigozhin}, {Ramirez}, {Reid},
  {Remillard}, {Rogstad}, {Rosecrans}, {Rowe}, {Sager}, {Sanders}, {Savadkin},
  {Saylor}, {Schaeffer}, {Schweiss}, {Semper}, {Serlemitsos}, {Shackelford},
  {Soong}, {Struebel}, {Vezie}, {Villasenor}, {Winternitz}, {Wofford},
  {Wright}, {Yang}, \& {Yu}}]{Gendreau16}
{Gendreau}, K.~C., {Arzoumanian}, Z., {Adkins}, P.~W., {et~al.} 2016, SPIE,
  9905, 1H, \dodoi{10.1117/12.2231304}

\bibitem[{{Ghisellini} {et~al.}(1988){Ghisellini}, {Guilbert}, \&
  {Svensson}}]{GGS88}
{Ghisellini}, G., {Guilbert}, P.~W., \& {Svensson}, R. 1988, \apjl, 334, L5,
  \dodoi{10.1086/185300}

\bibitem[{{Gierli{\'n}ski} {et~al.}(1999){Gierli{\'n}ski}, {Zdziarski},
  {Poutanen}, {Coppi}, {Ebisawa}, \& {Johnson}}]{G99}
{Gierli{\'n}ski}, M., {Zdziarski}, A.~A., {Poutanen}, J., {et~al.} 1999,
  \mnras, 309, 496

\bibitem[{{Gould} \& {Schr{\'e}der}(1967)}]{GS67}
{Gould}, R.~J., \& {Schr{\'e}der}, G.~P. 1967, Physical Review, 155, 1404,
  \dodoi{10.1103/PhysRev.155.1404}

\bibitem[{{Harrison} {et~al.}(2013){Harrison}, {Craig}, {Christensen},
  {Hailey}, {Zhang}, {Boggs}, {Stern}, {Cook}, {Forster}, {Giommi},
  {Grefenstette}, {Kim}, {Kitaguchi}, {Koglin}, {Madsen}, {Mao}, {Miyasaka},
  {Mori}, {Perri}, {Pivovaroff}, {Puccetti}, {Rana}, {Westergaard}, {Willis},
  {Zoglauer}, {An}, {Bachetti}, {Barri{\`e}re}, {Bellm}, {Bhalerao},
  {Brejnholt}, {Fuerst}, {Liebe}, {Markwardt}, {Nynka}, {Vogel}, {Walton},
  {Wik}, {Alexander}, {Cominsky}, {Hornschemeier}, {Hornstrup}, {Kaspi},
  {Madejski}, {Matt}, {Molendi}, {Smith}, {Tomsick}, {Ajello}, {Ballantyne},
  {Balokovi{\'c}}, {Barret}, {Bauer}, {Blandford}, {Brandt}, {Brenneman},
  {Chiang}, {Chakrabarty}, {Chenevez}, {Comastri}, {Dufour}, {Elvis}, {Fabian},
  {Farrah}, {Fryer}, {Gotthelf}, {Grindlay}, {Helfand}, {Krivonos}, {Meier},
  {Miller}, {Natalucci}, {Ogle}, {Ofek}, {Ptak}, {Reynolds}, {Rigby},
  {Tagliaferri}, {Thorsett}, {Treister}, \& {Urry}}]{Harrison13}
{Harrison}, F.~A., {Craig}, W.~W., {Christensen}, F.~E., {et~al.} 2013, \apj,
  770, 103, \dodoi{10.1088/0004-637X/770/2/103}

\bibitem[{{Kajava} {et~al.}(2019){Kajava}, {Motta}, {Sanna}, {Veledina}, {Del
  Santo}, \& {Segreto}}]{Kajava19}
{Kajava}, J.~J.~E., {Motta}, S.~E., {Sanna}, A., {et~al.} 2019, \mnras, 488,
  L18, \dodoi{10.1093/mnrasl/slz089}

\bibitem[{{Kawamuro} {et~al.}(2018){Kawamuro}, {Negoro}, {Yoneyama}, {Ueno},
  {Tomida}, {Ishikawa}, {Sugawara}, {Isobe}, {Shimomukai}, {Mihara},
  {Sugizaki}, {Nakahira}, {Iwakiri}, {Yatabe}, {Takao}, {Matsuoka}, {Kawai},
  {Sugita}, {Yoshii}, {Tachibana}, {Harita}, {Morita}, {Yoshida}, {Sakamoto},
  {Serino}, {Kawakubo}, {Kitaoka}, {Hashimoto}, {Tsunemi}, {Nakajima},
  {Kawase}, {Sakamaki}, {Maruyama}, {Ueda}, {Hori}, {Tanimoto}, {Oda},
  {Morita}, {Yamada}, {Tsuboi}, {Nakamura}, {Sasaki}, {Kawai}, {Sato},
  {Yamauchi}, {Hanyu}, {Hidaka}, {Yamaoka}, \& {Shidatsu}}]{Kawamuro18}
{Kawamuro}, T., {Negoro}, H., {Yoneyama}, T., {et~al.} 2018, Astron.\ Telegram,
  11399, 1

\bibitem[{{Lubi{\'n}ski}(2009)}]{Lubinski09}
{Lubi{\'n}ski}, P. 2009, \aap, 496, 557, \dodoi{10.1051/0004-6361:200810897}

\bibitem[{{Magdziarz} \& {Zdziarski}(1995)}]{MZ95}
{Magdziarz}, P., \& {Zdziarski}, A.~A. 1995, \mnras, 273, 837

\bibitem[{{Malzac} \& {Belmont}(2009)}]{MB09}
{Malzac}, J., \& {Belmont}, R. 2009, \mnras, 392, 570,
  \dodoi{10.1111/j.1365-2966.2008.14142.x}

\bibitem[{{McConnell} {et~al.}(2002){McConnell}, {Zdziarski}, {Bennett},
  {Bloemen}, {Collmar}, {Hermsen}, {Kuiper}, {Paciesas}, {Phlips}, {Poutanen},
  {Ryan}, {Sch{\"o}nfelder}, {Steinle}, \& {Strong}}]{McConnell02}
{McConnell}, M.~L., {Zdziarski}, A.~A., {Bennett}, K., {et~al.} 2002, \apj,
  572, 984, \dodoi{10.1086/340436}

\bibitem[{{Nied{\'z}wiecki} {et~al.}(2019){Nied{\'z}wiecki}, {Szanecki}, \&
  {Zdziarski}}]{Niedzwiecki19}
{Nied{\'z}wiecki}, A., {Szanecki}, M., \& {Zdziarski}, A.~A. 2019, \mnras, 485,
  2942, \dodoi{10.1093/mnras/stz487}

\bibitem[{{Poutanen} \& {Svensson}(1996)}]{PS96}
{Poutanen}, J., \& {Svensson}, R. 1996, \apj, 470, 249, \dodoi{10.1086/177865}

\bibitem[{{Poutanen} \& {Veledina}(2014)}]{PV14}
{Poutanen}, J., \& {Veledina}, A. 2014, \ssr, 183, 61,
  \dodoi{10.1007/s11214-013-0033-3}

\bibitem[{{Poutanen} \& {Vurm}(2009)}]{PV09}
{Poutanen}, J., \& {Vurm}, I. 2009, \apjl, 690, L97,
  \dodoi{10.1088/0004-637X/690/2/L97}

\bibitem[{{Roques} \& {Jourdain}(2019)}]{Roques19}
{Roques}, J.-P., \& {Jourdain}, E. 2019, \apj, 870, 92,
  \dodoi{10.3847/1538-4357/aaf1c9}

\bibitem[{{Roques} {et~al.}(2003){Roques}, {Schanne}, {von Kienlin},
  {Kn{\"o}dlseder}, {Briet}, {Bouchet}, {Paul}, {Boggs}, {Caraveo},
  {Cass{\'e}}, {Cordier}, {Diehl}, {Durouchoux}, {Jean}, {Leleux}, {Lichti},
  {Mandrou}, {Matteson}, {Sanchez}, {Sch{\"o}nfelder}, {Skinner}, {Strong},
  {Teegarden}, {Vedrenne}, {von Ballmoos}, \& {Wunderer}}]{Roques03}
{Roques}, J.~P., {Schanne}, S., {von Kienlin}, A., {et~al.} 2003, \aap, 411,
  L91, \dodoi{10.1051/0004-6361:20031501}

\bibitem[{{Sironi} \& {Beloborodov}(2020)}]{Sironi20}
{Sironi}, L., \& {Beloborodov}, A.~M. 2020, \apj, 899, 52,
  \dodoi{10.3847/1538-4357/aba622}

\bibitem[{{Sironi} {et~al.}(2016){Sironi}, {Giannios}, \&
  {Petropoulou}}]{Sironi16}
{Sironi}, L., {Giannios}, D., \& {Petropoulou}, M. 2016, \mnras, 462, 48,
  \dodoi{10.1093/mnras/stw1620}

\bibitem[{{Sironi} {et~al.}(2015){Sironi}, {Keshet}, \& {Lemoine}}]{Sironi15}
{Sironi}, L., {Keshet}, U., \& {Lemoine}, M. 2015, \ssr, 191, 519,
  \dodoi{10.1007/s11214-015-0181-8}

\bibitem[{{Stern} {et~al.}(1995){Stern}, {Begelman}, {Sikora}, \&
  {Svensson}}]{Stern95}
{Stern}, B.~E., {Begelman}, M.~C., {Sikora}, M., \& {Svensson}, R. 1995,
  \mnras, 272, 291

\bibitem[{{Sunyaev} \& {Titarchuk}(1980)}]{ST80}
{Sunyaev}, R.~A., \& {Titarchuk}, L.~G. 1980, \aap, 86, 121

\bibitem[{{Svensson}(1982)}]{Svensson82a}
{Svensson}, R. 1982, \apj, 258, 321, \dodoi{10.1086/160081}

\bibitem[{{Svensson}(1987)}]{Svensson87}
---. 1987, \mnras, 227, 403

\bibitem[{{Svensson}(1996)}]{Svensson96}
---. 1996, \aaps, 120, C475

\bibitem[{{Svensson} {et~al.}(1996){Svensson}, {Larsson}, \&
  {Poutanen}}]{SLP96}
{Svensson}, R., {Larsson}, S., \& {Poutanen}, J. 1996, \aaps, 120, C587

\bibitem[{{Torres} {et~al.}(2020){Torres}, {Casares}, {Jim{\'e}nez-Ibarra},
  {{\'A}lvarez-Hern{\'a}ndez}, {Mu{\~n}oz-Darias}, {Armas Padilla}, {Jonker},
  \& {Heida}}]{Torres20}
{Torres}, M.~A.~P., {Casares}, J., {Jim{\'e}nez-Ibarra}, F., {et~al.} 2020,
  \apjl, 893, L37, \dodoi{10.3847/2041-8213/ab863a}

\bibitem[{{Tucker} {et~al.}(2018){Tucker}, {Shappee}, {Holoien}, {Auchettl},
  {Strader}, {Stanek}, {Kochanek}, {Bahramian}, {ASAS-SN}, {Dong}, {Prieto},
  {Shields}, {Thompson}, {Beacom}, {Chomiuk}, {ATLAS}, {Denneau}, {Flewelling},
  {Heinze}, {Smith}, {Stalder}, {Tonry}, {Weiland}, {Rest}, {Huber}, {Rowan},
  \& {Dage}}]{Tucker18}
{Tucker}, M.~A., {Shappee}, B.~J., {Holoien}, T.~W.~S., {et~al.} 2018, \apjl,
  867, L9, \dodoi{10.3847/2041-8213/aae88a}

\bibitem[{{Ubertini} {et~al.}(2003){Ubertini}, {Lebrun}, {Di Cocco}, {Bazzano},
  {Bird}, {Broenstad}, {Goldwurm}, {La Rosa}, {Labanti}, {Laurent}, {Mirabel},
  {Quadrini}, {Ramsey}, {Reglero}, {Sabau}, {Sacco}, {Staubert}, {Vigroux},
  {Weisskopf}, \& {Zdziarski}}]{Ubertini03}
{Ubertini}, P., {Lebrun}, F., {Di Cocco}, G., {et~al.} 2003, \aap, 411, L131,
  \dodoi{10.1051/0004-6361:20031224}

\bibitem[{{Veledina} {et~al.}(2013){Veledina}, {Poutanen}, \& {Vurm}}]{VPV13}
{Veledina}, A., {Poutanen}, J., \& {Vurm}, I. 2013, \mnras, 430, 3196,
  \dodoi{10.1093/mnras/stt124}

\bibitem[{{Veledina} {et~al.}(2011){Veledina}, {Vurm}, \& {Poutanen}}]{VVP11}
{Veledina}, A., {Vurm}, I., \& {Poutanen}, J. 2011, \mnras, 414, 3330,
  \dodoi{10.1111/j.1365-2966.2011.18635.x}

\bibitem[{{Walter} \& {Xu}(2017)}]{Walter17}
{Walter}, R., \& {Xu}, M. 2017, \aap, 603, A8,
  \dodoi{10.1051/0004-6361/201629347}

\bibitem[{{Wang} {et~al.}(2020){Wang}, {Ji}, {Zhang}, {M{\'e}ndez}, {Qu},
  {Maggi}, {Ge}, {Qiao}, {Tao}, {Zhang}, {Altamirano}, {Zhang}, {Ma}, {Lu},
  {Li}, {Huang}, {Zheng}, {Chen}, {Chang}, {Tuo}, {G{\"u}ng{\"o}r}, {Song},
  {Xu}, {Cao}, {Chen}, {Liu}, {Bu}, {Cai}, {Chen}, {Chen}, {Chen}, {Chen},
  {Cui}, {Cui}, {Deng}, {Dong}, {Du}, {Fu}, {Gao}, {Gao}, {Gao}, {Gu}, {Guan},
  {Guo}, {Han}, {Huo}, {Jia}, {Jiang}, {Jiang}, {Jin}, {Jin}, {Kong}, {Li},
  {Li}, {Li}, {Li}, {Li}, {Li}, {Li}, {Li}, {Li}, {Li}, {Liang}, {Liao}, {Liu},
  {Liu}, {Liu}, {Liu}, {Lu}, {Lu}, {Luo}, {Luo}, {Meng}, {Nang}, {Nie}, {Ou},
  {Sai}, {Shang}, {Song}, {Sun}, {Tan}, {Wang}, {Wang}, {Wang}, {Wang}, {Wang},
  {Wen}, {Wu}, {Wu}, {Wu}, {Xiao}, {Xiao}, {Xiong}, {Yang}, {Yang}, {Yang},
  {Yang}, {Yi}, {Yin}, {You}, {Zhang}, {Zhang}, {Zhang}, {Zhang}, {Zhang},
  {Zhang}, {Zhang}, {Zhang}, {Zhang}, {Zhang}, {Zhang}, {Zhang}, {Zhang},
  {Zhang}, {Zhang}, {Zhang}, {Zhao}, {Zhao}, {Zhou}, {Zhou}, {Zhuang}, {Zhu},
  {Zhu}, \& {Wang}}]{Wang20_HXMT}
{Wang}, Y., {Ji}, L., {Zhang}, S.~N., {et~al.} 2020, \apj, 896, 33,
  \dodoi{10.3847/1538-4357/ab8db4}

\bibitem[{{Wardzi{\'n}ski} \& {Zdziarski}(2001)}]{WZ01}
{Wardzi{\'n}ski}, G., \& {Zdziarski}, A.~A. 2001, \mnras, 325, 963,
  \dodoi{10.1046/j.1365-8711.2001.04387.x}

\bibitem[{{Wardzi{\'n}ski} {et~al.}(2002){Wardzi{\'n}ski}, {Zdziarski},
  {Gierli{\'n}ski}, {Grove}, {Jahoda}, \& {Johnson}}]{Wardzinski02}
{Wardzi{\'n}ski}, G., {Zdziarski}, A.~A., {Gierli{\'n}ski}, M., {et~al.} 2002,
  \mnras, 337, 829, \dodoi{10.1046/j.1365-8711.2002.05914.x}

\bibitem[{{Wilms} {et~al.}(2000){Wilms}, {Allen}, \& {McCray}}]{WAMC00}
{Wilms}, J., {Allen}, A., \& {McCray}, R. 2000, \apj, 542, 914,
  \dodoi{10.1086/317016}

\bibitem[{{Yuan} \& {Narayan}(2014)}]{YN14}
{Yuan}, F., \& {Narayan}, R. 2014, \araa, 52, 529,
  \dodoi{10.1146/annurev-astro-082812-141003}

\bibitem[{{Zdziarski}(1985)}]{Zdziarski85}
{Zdziarski}, A.~A. 1985, \apj, 289, 514, \dodoi{10.1086/162912}

\bibitem[{{Zdziarski}(1986)}]{Z86}
---. 1986, \apj, 303, 94, \dodoi{10.1086/164055}

\bibitem[{{Zdziarski}(1988)}]{Zdziarski88}
---. 1988, \apj, 335, 786, \dodoi{10.1086/166967}

\bibitem[{{Zdziarski} {et~al.}(2021{\natexlab{a}}){Zdziarski}, {De Marco},
  {Szanecki}, {Nied{\'z}wiecki}, \& {Markowitz}}]{Zdziarski21}
{Zdziarski}, A.~A., {De Marco}, B., {Szanecki}, M., {Nied{\'z}wiecki}, A., \&
  {Markowitz}, A. 2021{\natexlab{a}}, \apj, 906, 69,
  \dodoi{10.3847/1538-4357/abca9c}

\bibitem[{{Zdziarski} {et~al.}(2021{\natexlab{b}}){Zdziarski}, {Dzie{\l}ak},
  {De Marco}, {Szanecki}, \& {Nied{\'z}wiecki}}]{Zdziarski21b}
{Zdziarski}, A.~A., {Dzie{\l}ak}, M.~A., {De Marco}, B., {Szanecki}, M., \&
  {Nied{\'z}wiecki}, A. 2021{\natexlab{b}}, \apjl, 909, L9,
  \dodoi{10.3847/2041-8213/abe7ef}

\bibitem[{{Zdziarski} {et~al.}(1993){Zdziarski}, {Lightman}, \&
  {Maciolek-Nied{\'z}wiecki}}]{ZLM93}
{Zdziarski}, A.~A., {Lightman}, A.~P., \& {Maciolek-Nied{\'z}wiecki}, A. 1993,
  \apjl, 414, L93, \dodoi{10.1086/187004}

\bibitem[{{Zdziarski} {et~al.}(2017){Zdziarski}, {Malyshev}, {Chernyakova}, \&
  {Pooley}}]{ZMC17}
{Zdziarski}, A.~A., {Malyshev}, D., {Chernyakova}, M., \& {Pooley}, G.~G. 2017,
  \mnras, 471, 3657.
\newblock \doarXiv{1607.05059}

\bibitem[{{Zdziarski} {et~al.}(2020){Zdziarski}, {Szanecki}, {Poutanen},
  {Gierli{\'n}ski}, \& {Biernacki}}]{Z20_thcomp}
{Zdziarski}, A.~A., {Szanecki}, M., {Poutanen}, J., {Gierli{\'n}ski}, M., \&
  {Biernacki}, P. 2020, \mnras, 492, 5234, \dodoi{10.1093/mnras/staa159}

\end{thebibliography}
\bibliographystyle{aasjournal}

\label{lastpage}
\end{document}